\documentclass[12pt]{article}
\usepackage{epsfig}
\usepackage{amssymb,amsmath}
\usepackage{mathdots}                %
\usepackage{cite}
\usepackage{graphics}
\usepackage{hyperref}
\usepackage{bbm}
\usepackage{multirow}
\usepackage{booktabs}
\usepackage{tikz}
\usepackage{float}
\restylefloat{table}
\tikzset{epsilon/.style={circle,draw,fill=white,minimum size=8pt,inner sep=0pt,font=\scriptsize}}

         \let\d=d

       \let\C=\Chi    
\newcommand{\be}{\begin{equation}}
\newcommand{\ee}{\end{equation}}
\newcommand{\bea}{\begin{eqnarray}}
\newcommand{\eea}{\end{eqnarray}}
\newcommand{\ba}{\begin{array}}
\newcommand{\ea}{\end{array}}
\def\nn{\nonumber}

\newcommand{\ft}[2]{{\textstyle\frac{#1}{#2}}}
\newcommand{\Z}{{\mathbb Z}}
\newcommand{\R}{{\mathbb R}}
\newcommand{\C}{{\mathbb C}}

\def\calc         {{\cal C}}

\def\calm         {{\cal M}}
\def\caln         {{\cal N}}

\def\delbar       {\bar\partial}
\def\ii           {{\rm i}}

\begin{document}

\font\cmss=cmss10 \font\cmsss=cmss10 at 7pt

\vskip -0.5cm
\rightline{\small{\tt ROM2F/2013/06}}
\rightline{\small{\tt KCL-mth-13-07}}

\vskip .7 cm
%\hfill IC/2004/ \vskip .1in \hfill CPHT \vskip .1in \hfill hep-th/yymmnnn

\hfill
\vspace{18pt}
\begin{center}
{\Large \textbf{U-folds as K3 fibrations}}
\end{center}

%\vspace{12pt}
%

%\vspace{12pt}
%
\vspace{6pt}
\begin{center}

  {\textsl{ Andreas P. Braun $^\dagger$ \footnote{\scriptsize \tt andreas.braun@kcl.ac.uk},
    Francesco Fucito $^{\ddagger}$ \footnote{\scriptsize \tt fucito@roma2.infn.it} and
     Jose Francisco Morales $^{\ddagger }$ \footnote{\scriptsize \tt francisco.morales@roma2.infn.it}  }}

\vspace{1cm}

 $ \dagger $ \textit{\small King's College, Department of Mathematics \\
Strand, London, WC2R 2LS, UK    }\\  \vspace{6pt}

$\ddagger$ \textit{\small I.N.F.N. Sezione di Roma ``TorVergata''  \\  Dipartimento di Fisica, Universit\`a di Roma ``TorVergata", \\
Via della Ricerca Scientica, 00133 Roma, Italy }\\  \vspace{6pt}

\end{center}

\begin{center}
\textbf{Abstract}
\end{center}
We study ${\cal N}=2$ four-dimensional flux vacua describing intrinsic non-perturbative systems of 3 and 7 branes in type IIB string theory.
The solutions are described as compactifications of a G(ravity) theory on a Calabi Yau threefold which consists of a fibration of
an auxiliary K3 surface over an $S^2$ base. In the spirit of F-theory, the complex structure of the K3 surface varying over the base codifies the details of the fluxes,
the dilaton and the warp factors in type IIB string theory. We discuss in detail some simple examples of geometric and non-geometric solutions where
the precise flux/geometry dictionary  can be explicitly worked out. In particular, we describe non-geometric T-fold solutions exhibiting non-trivial
T-duality monodromies exchanging 3- and 7-branes.

\vspace{4pt} {\small

\noindent }

\numberwithin{equation}{section}
\newpage
\tableofcontents

\section{Introduction}
F-Theory \cite{Vafa:1996xn,Sen:1996vd} describes fully non-perturbative solutions of type IIB theory (with a varying axio-dilaton field)
in purely geometric terms.  It links type IIB to M-Theory and elucidates the geometric
origin of the $SL(2,\mathbb{Z})$ S-duality symmetry of type IIB theory. Besides the conceptual beauty, F-Theory is a powerful tool to build and analyse
semi realistic gauge theories (F-theory GUT's) with non-trivial strong coupling dynamics  \cite{Beasley:2008dc,Beasley:2008kw,Donagi:2008ca,Hayashi:2008ba}
and provides explicit D-brane set-ups where the gauge-gravity correspondence can be tested at the non-perturbative level \cite{Billo:2011uc,Fucito:2011kb,Billo:2012st}.

In this paper we present a construction which, in a similar spirit, makes use of the larger U-duality group of type IIB string theory
compactified on a $K3$ surface. Type IIB supergravity compactified on a K3 surface is described by an effective ${\cal N}=(2,0)$ six-dimensional supergravity with 105 scalars and U-duality group $SO(5,21,\Z)$.
A class of supersymmetric solutions (vacua) preserving ${\cal N}=2$ supersymmetries in this theory can be found by allowing a subset of the scalars
spanning the moduli space submanifold
 \be\label{truncmsn}
\calm_{\rm BPS} ={\rm O}(\Gamma_{2,18})\backslash { {\rm O}(2,18;\mathbb{R})\over
 {\rm O}(2;\mathbb{R}) \times {\rm O}(18;\mathbb{R})}
\ee
 to vary holomorphically on a complex plane \cite{Martucci:2012jk}.
The scalars $\varphi^I=(\tau,\sigma,\beta^a)$, $a=1,..16$, spanning (\ref{truncmsn}) follow from the reduction along K3 of
the axio-dilaton field, the four form, the warp factor and the NS-NS/R-R two form potentials, respectively. The solutions, that
we dub U-folds, are specified by a set of holomorphic functions $\varphi^I(z)$ defined on a
punctured complex plane (described by the coordinate $z$) up to non-trivial monodromies of the U-duality group
${\rm O}(\Gamma_{2,18}) \sim {\rm O}(2,18;\mathbb{Z})$.

In F-theory the positions of the punctures specify the locations of the 7-branes and
the $SL(2,\mathbb{Z})$ monodromies give their $(p,q)$ charges.
 Analogously the location of punctures and monodromies in the U-fold
solution encode the positions and type of a richer set of exotic branes  in the ${\rm O}(2,18;\mathbb{Z})$
U-duality orbit of a D7- or a D3-brane (see \cite{deBoer:2012ma} for a recent discussion of exotic branes in string theory).
Indeed, in the same way that $\tau$ is sourced by D7-branes,  $\sigma$ is sourced by D3-branes and therefore the general U-fold solutions  describe systems of 3- and 7-branes  non-perturbatively completed by brane instantons. We remark that at generic points in the moduli space, the solutions are intrinsically non-perturbative so that only under very special circumstances one can give a perturbative D-brane description (see  \cite{Martucci:2012jk} for explicit choices of fibrations  realizing systems of fractional D3 and D7 brane in type IIB theory).
Like in F-theory, the presence of branes curves the plane $\C$ and when a maximal number is reached
the plane is compactified to a sphere $S^2$  \cite{Greene:1989ya}. Here we limit ourselves to this compact case since we are interested in
four-dimensional flux vacua.

It is important to observe that the space in which the scalars in the U-fold solution live, (\ref{truncmsn}), is isomorphic to the moduli space
of complex deformations of a K3 surfaces of elliptic type. This is not a coincidence. There is, in fact, an alternative way of building an ${\cal N}=2$
vacuum by simply compactifying type IIB theory on a Calabi-Yau threefold given by a fibration of a K3 surface over $S^2$ with no fluxes.
The U-duality group $SO(5,21,\Z)$ of type IIB string theory on a K3 surface relates the two solutions, translating the geometry of the threefold
into fluxes and vice versa. Indeed a point in the moduli space of compactifications of type IIB string theory on a K3 surface is specified by
fixing a positive definite 5-plane $\Sigma_5$ in a $26$-dimensional space with signature $(5,21)$. The orientation of this
5-plane determines the metric as well as the fluxes. In this language, both a K3 fibred Calabi-Yau threefold and the U-fold solution of
\cite{Martucci:2012jk} can be described by letting two of the five vectors spanning $\Sigma_5$ vary within a subspace of signature $(2,18)$. The orientation
of the 2-plane is described by the Grassmannian (\ref{truncmsn}).
It is a pure matter of convention  to identify the 2-plane with that specifying the geometry or fluxes.

\begin{figure}[!h]
\scalebox{0.5}{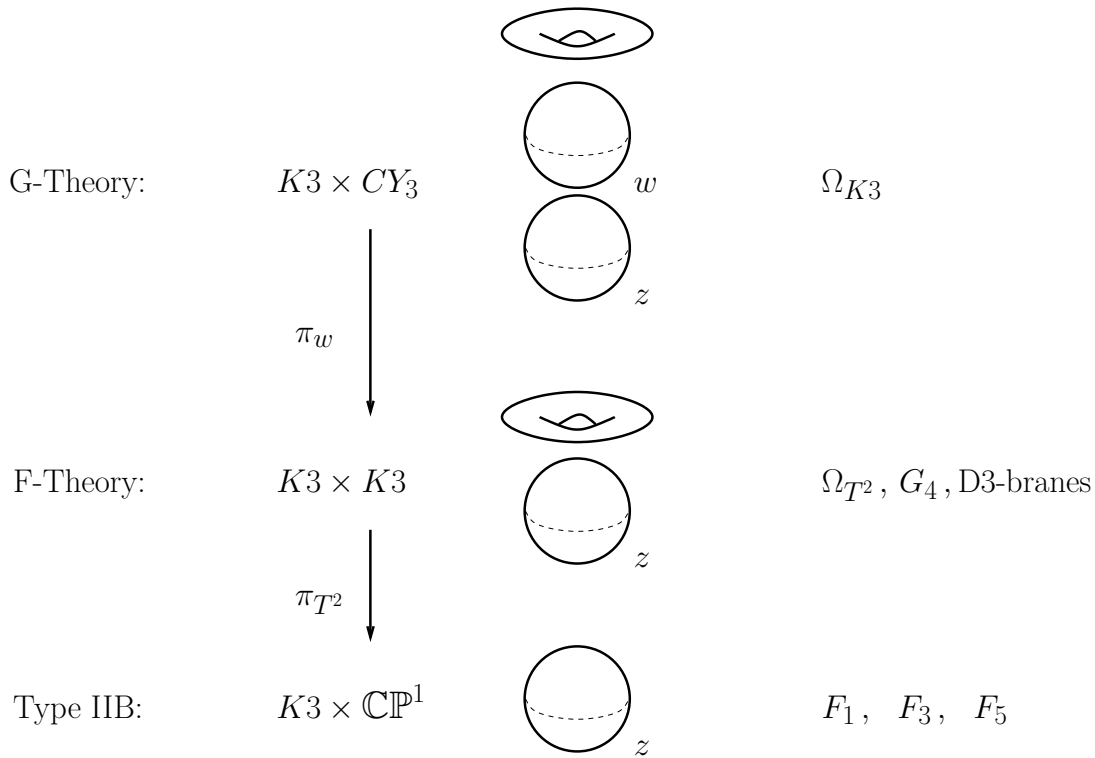}
\caption{G-theory lift. Coordinates $z$ and $w$ runs along the spacetime and the base of the auxiliary elliptic K3 surface  respectively.
\label{IIBFGfigure}}
\end{figure}

In the spirit of F-theory, one can  view the threefold (a K3 fibration over $S^2$) underlying the dual geometric solution as the compactification
manifold of a new theory, we dub as {\it G-theory} (with G standing for Gravity, or Geometry or Grandfather). Like in F-theory, the geometry of
the threefold codifies the details of the flux solution. In particular, the holomorphic functions $\varphi^I(z)$ describing how the complex
structure of the K3 surface vary over the base will describe how   the axio-dilaton field $\tau(z)$, the warp factor/four-form field $\sigma(z)$
and the two-form NS-NS and R-R potentials $\beta^a(z)$ vary over $S^2$. The degenerations of the K3 fibre signal the presence of spacetime branes.
Alternatively one can think of G-theory as a purely geometric lift of F-theory (or M-theory) with non-trivial
$G_4$-form fluxes \cite{Becker:1996gj,Dasgupta:1999ss} and fivebrane charges. In
figure \ref{IIBFGfigure} we display the general structure of the G-theory lift.  We remark that, as in the case of F-theory, only the complex structure of the K3 fibre
has a meaning in the realm of G-theory. The K\"ahler structure moduli can be thought of as being frozen to zero.

Alternatively, one can think of
the threefold as an elliptic fibration over a Hirzebruch surface $F_n$ (a $S^2$ fibered over $S^2$ with winding number $n$).
These geometries have been extensively studied in the past in the context of F-theory/heterotic duality in six-dimension.
Indeed, the space (\ref{truncmsn}) is also the moduli space of heterotic string theory on $T^2$. In this paper, we exploit the detailed knowledge of
these geometries coming from the F-theory experience to produce U-fold solutions and explain how the 3- and 7-brane content of the solutions
is codified in the threefold geometry. We discuss a number of examples where the explicit dictionary between
geometry and fluxes can be worked out in detail. In particular, we find examples of non-geometric U-fold solutions exhibiting non-trivial
T-duality monodromies exchanging 3- and 7-branes\footnote{J.F.M. thanks D. Waldram  for an interesting discussion on this point.}. These
provide one of the simplest explicit realizations of T-folds in string theory\footnote{Stringy set ups where D3- and D7-branes are identified
have been realized in the past in terms of asymmetric orbifolds  involving the quotient by the T-duality action on all four
coordinates \cite{Bianchi:1999uq}. }.
(see \cite{Hellerman:2002ax,Hull:2004in,Flournoy:2004vn,Dabholkar:2005ve,Gray:2005ea,Vegh:2008jn,Grana:2008yw,McOrist:2010jw}
for previous works on T-folds). A related construction of string vacua where fluxes are codified into geometry has been
recently developed in \cite{Hohm:2011zr,Hohm:2011dv,Coimbra:2011nw,Coimbra:2011ky,Coimbra:2012af}. This approach has been
pioneered in \cite{Siegel:1993xq,Siegel:1993th}, see also \cite{West:2010ev,Rocen:2010bk} for an embedding of these ideas into a
broader perspective. A more complete list of references can be found in \cite{Andriot:2011uh,Andriot:2013txa,Aldazabal:2013sca}.

The plan of the paper is the following: In Section \ref{sechol} we review the construction of U-fold solutions in ${\cal N}=(2,0)$
supergravity and describe the details of the underlying threefold geometry. In Section  \ref{sexamples} we discuss three examples of U-fold solutions associated to K3 fibrations with two or three active complex structure moduli. These geometries are very well studied in the context of F-theory/heterotic duality and are associated to K3 fibres with singularities
of type $E_8\times E_8$, $E_8 \times E_7$ and $D_4^4$ respectively. Appendix \ref{sk3} collects some background material and details on elliptic K3 surfaces.

\section{U-fold solutions}

\subsection{The supergravity solution}
\label{sechol}

Here we review the U-fold construction in \cite{Martucci:2012jk} of $\caln=2$ supersymmetric solutions of type IIB supergravity (see \cite{Grana:2001xn} for local
versions of these solutions).
We consider type IIB supergravity on a warped space with topology $\mathbb{R}^{1,3}\times Y$
 a varying axio-dilaton field and non-trivial NS-NS  and R-R fluxes. The ten-dimensional metric
is given in the Einstein frame by
 \be\label{10dmetric}
 \d s^2_{\rm E} =  e^{2 A}\d x^\mu\d x_\mu+ e^{-2A} ds^2_Y
  \ee
  with $A$ a warp factor,
  \be
\d s^2_Y=   \left( e^{-\phi}
 |h(z)|^2 \d z \d\bar z +\d s^2_{X} \right)
  \ee
and  $\d s^2_X$  the Ricci flat metric on a K3 surface $X$.
  Here $z$ is a coordinate on the complex plane $\C$ with punctures at the positions of branes,
  $h(z)$ is a holomorphic function (away from the punctures) and $\phi$ is the dilaton field.
  The surface $Y$ admits complex, $\Omega$, and K\"ahler, $J$, structures
  \be
  \Omega=h(z) dz\wedge w   \qquad     J=j-\ft{i}{2} e^{-\phi}
 |h(z)|^2 \d z \wedge  \d\bar z
  \ee
  satisfying the Calabi-Yau conditions
  \be
      d\Omega=dJ=0
  \ee
   Here $\Omega$ and $J$ are the complex and K\"ahler structures of the K3 surface.  As we explained before, the space $Y$ can  be compact or non compact depending on the number of branes but here we limit ourselves to the compact case
    where the number of branes is maximal and $\C$ is compactified to a $S^2$.

  NS-NS and R-R potentials are
 conveniently packed into a set of complex functions
  \be
  \begin{aligned}
   \tau (z)&=    C_0+\ii e^{-\phi } \\
 \sigma (z)&=     \int_{X} \left(C_4-\ft{\ii}{2} e^{-4A}  J\wedge J+B\wedge C_2+\ft12 \tau B \wedge B \right) \\
 % \int_{X} \left( c_4+\ii \, e^{\phi-4A}
  % +  c\cdot   b +  \ft12   \tau  \,b \cdot  b \right)
   \beta_a (z)& = \int_{\calc_a}  (C_2+\tau B)    \label{sols0}
 \end{aligned}
 \ee
 varying holomorphically on $z$ (away from punctures), i.e.
\be\label{holcond0}
\delbar\tau=0\, ,\quad \delbar\sigma=0\, ,\quad \delbar\beta^a=0
\ee
The solutions  $\varphi^I=(\tau,\sigma,\beta^a)$, parametrize the moduli space (\ref{truncmsn}).
$\Gamma_{2,18} \approx \Gamma_{2,2}\oplus \Gamma_{16}$ is the
lattice orthogonal to the three plane $\Sigma\in \Gamma_{5,21}$ defined by a choice of a hyper K\"ahler structure $(J,\Omega,\bar\Omega)$ on $K3$.
The cycles $\calc_a$ in
(\ref{sols0})  are associated to $\Gamma_{16}$  with intersection
matrix  $\Delta_{ab}=\int_X [\calc_a] \wedge [\calc_b]$ given in terms of the Poincar\'e dual forms $[\calc_a]$.

 \begin{table}[t]
 \centering
\begin{tabular}{|c|c|}
\hline
generator & non-trivial action\\
 \hline\hline
 $S$ & $\tau\rightarrow -\frac{1}{\tau} \qquad~  \sigma\rightarrow \sigma-\frac{1}{2\tau} \Delta_{ab} \beta^a\,\beta^b      \qquad~ \beta^a\rightarrow \frac{1}{\tau}\beta^a$ \\
  $T$ & $\tau\rightarrow \tau+1$\\
 $W_{a}$ & $\quad \beta^b\rightarrow \beta^b+\delta^b_a $\\
 $ R $ & $\tau\leftrightarrow \sigma$\\
 \hline
\end{tabular}
\caption{Generators of the $SO(2,2+n,\Z)$ U-duality group}
\label{tab:duality1}
\end{table}
 We summarize in Table \ref{tab:duality1} the action of the U-duality group
 ${\rm O}(\Gamma_{2,18}) \sim {\rm O}(2,18;\mathbb{Z})$  on the fields $\varphi^I$.
  U-fold solutions are defined in terms of a set of holomorphic functions $\varphi^I(z)$ on the (compactified) complex plane $\C$ with
  non-trivial U-duality monodromies around the punctures. As we explained in the introduction, the functions $\varphi^I(z)$
  can be codified by the geometry of a Calabi-Yau threefold given by an elliptic K3 surface fibered over the complex plane, $\C$ (or its compactification, $S^2$).
     In the following we describe the details of these geometries and the flux/geometry dictionary.

\subsection{U-folds as K3 fibrations }\label{weierk3fibred}

 The geometry we have in mind is displayed in figure \ref{IIBFGfigure}. It is made out of an auxiliary K3 manifold fibered over the
space $S^2$ (or $\mathbb{CP}^1$) in such a way as to produce a Calabi-Yau threefold.
 As we want the K3 fibre to be elliptic, this Calabi-Yau threefold can also be viewed as an elliptic fibration over a Hirzebruch surface, $\mathbb{F}_n$, made out of the sphere, $S^2$, and the base of K3 (another $S^2$).
As the axio-dilaton field is part of the complex structure moduli of the K3 fibre, F-theory compactifications will appear as
a special subset of our solutions, where only the complex structure of the torus fibre inside the auxiliary K3 varies over the spacetime sphere.

 The Calabi-Yau threefold  can be described as a hypersurface (given by a homogeneous equation $\Sigma$) in a toric ambient
 space $\left(\C^7- Z\right)/\C^{*3}$ described by the weight system
\begin{equation}
 \begin{array}{c|ccccccc}
 \Sigma & y  & x & u & w_0 & w_1 & z_0 & z_1\\
  \hline
  6& 3 & 2 & 1 & 0 & 0 & 0 & 0 \\
  12 & 6 & 4 & 0 & 1 & 1 & 0 & 0 \\
 12+6n & 6+3n & 4+2n & 0 & 0 & n & 1 & 1
 \end{array}\, .
 \label{weights}
\end{equation}
 It is simply given by the  Weierstrass equation
\begin{equation}\label{weierfibredoverp1}
 \Sigma: \qquad y^2=x^3+x \, f \, u^4+g \,  u^6 ,
\end{equation}
with
\bea
f=f_{8,8+4n}(\vec w,\vec z)   \qquad ~~~~~~~~~  g=g_{12,12+6n}(\vec w,\vec z)
\eea
polynomials of the indicated degrees with respect to the last two $\C^*$-actions in  (\ref{weights}).  The Weierstrass equation (\ref{weierfibredoverp1}) defines a torus at each point $u,\vec w,\vec z$.
The coordinates $\vec w,\vec z$ span two spheres fibered with winding number $n$, i.e. a Hirzebruch surface $F_n$. When we are in a patch where
$u,w_0,z_0$ are non-vanishing, the $\C^*$ identifications may be used to fix these coordinates to $1$. We will always work in this chart, so that
$w_1=w$, $z_1=z$ become affine coordinates on $F_n$ while the $(x,y)$ coordinates satisfying (\ref{weierfibredoverp1}) describe a torus.

Alternatively one can think of the threefold as a K3 surface fibered over a base $S^2$ parametrized by $z$. Over any point
$z$ in the
base of this fibration the complex structure of the K3 is determined by $18$ complex parameters. This can
be seen from (\ref{weierfibredoverp1}) as follows. For a fixed $z$, $f$ and $g$ are polynomials in $w$ of degree $8$ and $12$, respectively.
These polynomials have $9+13=22$ coefficients. We may use the $SL(2,\C)$ and the rescaling symmetry to fix $4$ of these coefficients to arbitrary values,
so that only $18$ independent moduli remain. In the Calabi-Yau threefold, these $18$ complex parameters are given by the holomorphic functions  $\varphi^I(z)$ varying over the base. We identify these functions with the profiles
of the NS-NS, R-R forms, warp factors and axio-dilaton, i.e. $\varphi^I= (\sigma(z),\tau(z),\beta_a(z))$.

The functions $\varphi^I(z)$ are determined by the periods of the holomorphic two form
\begin{equation}\label{omega2}
\Omega =  \frac{dx\wedge dw}{\sqrt{x^3+f x +g}} \, .
\end{equation}

In practice, an explicit evaluation of $\varphi^I(z)$ is  in general technically involved, so that we limit our discussion
to specific choices of $f$ and $g$ in the following. Still even for simple choices of $f$ and $g$, the physics of the corresponding
U-fold solution is very rich and illustrates already the main features of the general solution.
In particular, special choices of $f$ and $g$ correspond to a restriction to a smaller number of active moduli and therefore to a
subgroup of the U-duality group.

In the following we will discuss in details some simple examples of Calabi-Yau threefolds where the moduli/flux dictionary can be worked out quite explicitly.

\section{Examples}
\label{sexamples}

The elliptic K3 surfaces we introduce are characterized by (blow ups of) singularities of higher rank. We have collected some background on
elliptic K3 surfaces in the appendix.
We recall that the lattice of integral two-cycles of a K3 surface,
  $\Gamma^{3,19} \equiv H^2(K3,\Z)$, can be always written as
 \be
 \Gamma^{3,19} = U\oplus U\oplus U \oplus E_8 \oplus E_8   \label{u3e82}
 \ee
  with $U$ a two dimensional lattice with intersection matrix $(^{01}_{10})$. The intersection
  $ \mbox{Pic} \equiv H^{1,1}(K3)\cap H^2(K3,\Z)$ is known as the
    \emph{Picard lattice} of the K3 surface. For elliptic K3 surfaces with a section (the case we have in mind) the Picard lattice always contains
    the $U$ lattice (associated to the base and the fibre) and the lattice describing the blow up cycles at the singularity.
    The orthogonal complement of the Picard lattice is known as the \emph{transcendental lattice} of the K3 surface and will be denoted by $T_X$.
    It gives rise to the space of complex deformations of the K3 surface which leave the Picard lattice fixed. These moduli will be identified with
    the active six-dimensional scalars in the U-fold solution.

  \begin{table}[h]
  \centering
\begin{tabular}{|c|c|c|}
\hline
 \#~{\rm moduli} & {\rm Picard~lattice}  &   {\rm Transcendental~ lattice}\\
 \hline
  2 &  $U\oplus E_8 \oplus E_8$   &   $U\oplus U$\\
  3 &  $U\oplus E_8\oplus E_7$   &   $U\oplus U\oplus (-2)$ \\
 2  & $U\oplus D_4^{\oplus 4}$  & $U(2)\oplus U(2)$   \\
 \hline
\end{tabular}
\caption{The examples of K3 surfaces with few complex structure deformations considered.  Here, $U(2)$ denotes a lattice with
intersection matrix $(^{02}_{20})$ and $(n)$ denotes a one-dimensional lattice generated by a vector squaring to $n$. }
 \label{tablelattices}
\end{table}
Table \ref{tablelattices} summarizes the main three working examples. The relevant geometries are discussed in detail in the appendix.
  The first and third cases are parametrized by two complex structure moduli  denoted $(\tau,\sigma)$
  that will be identified with the axio-dilaton and warp/four-form fields in type IIB theory. The second example includes also a
non-trivial NS-NS/R-R three form flux.

  In the case with two moduli, the space of the complex structure deformations of the K3 surface $X$ is
  \be
  {\cal M}= {\rm O}(T_X)\backslash { {\rm O}(2,2;\mathbb{R})\over
 {\rm O}(2;\mathbb{R}) \times {\rm O}(2;\mathbb{R})}\approx   {\rm O}(T_X)\backslash \left( {SL(2,\R)\over U(1) } \right)^2 \, .
 \label{moduli22}
  \ee
  with ${\rm O}(T_X)$ a subgroup of the U-duality group
  \be
 {\rm O}(\Gamma_{2,2})=SO(2,2,\Z)\approx  \Z_2 \times SL(2,\Z)_{\tau} \times
   SL(2,\Z)_{\sigma}  \label{gamma22}
\ee
 The $\Z_2$ acts by permuting the two factors and it is the generator $R$ in Table \ref{tab:duality1}.
 Besides this $\Z_2$ action, the space (\ref{moduli22}) describes the
   complex structures of two factorized tori. In the case for which
   $ \mbox{Pic}=U\oplus D_4^{\oplus 4} $, we will see that the $\Z_2$ holonomies are trivial. Therefore one can
    think of this fibration  as a double elliptic fibration where at each point $z$ we have
 two factorized tori with complex structures $\sigma$ and $\tau$.
On the other hand for $\mbox{Pic}=U\oplus E_8\oplus E_8 $, there are non-trivial $\Z_2$ monodromies around the singularities in the z-plane and
therefore the 3- and 7-branes are consistently identified. Indeed, going around these points the 3-branes are exchanged with the 7-branes and the
solution gets back to itself up to an action of the T-duality element $R$ of Table \ref{tab:duality1}. As we discuss in detail later, the distinction
between these two cases can be traced back to the different transcendental lattices of the corresponding K3 surfaces.

In the case of a K3 surface $X$ with three moduli the space of complex structure deformations is
\be
  {\cal M}= {\rm O}(T_X)\backslash { {\rm O}(2,3;\mathbb{R})\over
 {\rm O}(2;\mathbb{R}) \times {\rm O}(3;\mathbb{R})} \label{moduli23} \, .
   \ee
 with ${\rm O}(T_X)$ a subgroup of the U-duality group
 \be
 {\rm O}(\Gamma_{2,3})=SO(2,3,\Z)\sim   Sp(4,\Z) \, .
     \label{gamma23}
\ee
Notice that $Sp(4,\Z)$ is also the modular group of a genus two Riemann surface with period matrix $\Omega={\left(^{\sigma \beta}_{\beta \sigma} \right)}$.
In these cases, the elliptic fibration can be thought of as a fibration of a genus two  Riemann surface over $S^2$ \cite{Martucci:2012jk}.
The threefold now describes a system containing both regular and fractional 3- and 7-branes or equivalently non-trivial
NS-NS $H_3$ and R-R $F_{1,3,5}$  fluxes. Again non trivial $R$-monodromies will occur
signalling the non-geometric nature of the U-fold solution.

\subsection{$E_8\times E_{8}$ fibration: two moduli}

Let us consider the following Calabi-Yau hypersurface defined by the Weierstrass equation (\ref{weierfibredoverp1}) with\footnote{In the
stable degeneration limit ($b_{12}$ and $a_8$ going to infinity keeping $a^3/b^2$ fixed), this geometry describes
the F-Theory dual of the heterotic string compactified on $T^2$ in the absence of Wilson lines and at large volume \cite{Morrison:1996pp}.
A fibration of this K3 manifold away from the stable degeneration limit has been used in \cite{McOrist:2010jw} to give a F-theory description
of a non-geometric heterotic compactification.}
   \bea\label{K3e8e8eq}
 f &=&  a_8(z) \, w^4  \nn\\
 g &=& d_{12- n}(z) w^7 + b_{12} (z)\, w^6+ d'_{12+ n}(z) w^5 \, .
 \eea
Here, the subscripts denote the degree of the polynomials in the variable $z$. The discriminant is $\Delta=w^{10} P_4(w)$
with $P_4(w)$ a polynomial of order four in $w$. This K3 surface is
singular at $w=0$ and $w=\infty$ with both singularities of type $E_8$ according to the Kodaira classification
(see Table \ref{kodaira} in the Appendix).
The moduli space of complex structure deformations of the K3 surface (\ref{moduli22}) can be characterized by \cite{kumarelkies,LopesCardoso:1996hq}
  \bea
   -{a^3\over 27\, d\, d'} &=&    j_1\, j_2 \nn\\
     {b^2\over 4\, d\, d'} &=&   ( j_1-1)( j_2-1) \label{j1j2}
    \eea
 with
      \be
j_i={j (\tau_i) \over 1728}  \qquad       \tau_i=(\sigma,\tau)
\ee
 and $j={E_4^3\over  \eta^{24}} $\footnote{ Here $E_4=  \ft12 (\vartheta_2^8+\vartheta_3^8+\vartheta_4^8)$ and
$E_6= \ft12 (\vartheta_2^4+\vartheta_3^4) (\vartheta_3^4+\vartheta_4^4)(\vartheta_4^4-\vartheta_2^4)$
satisfy the identity $E_4^3-E_6^2=1728\, \eta^{24}$.
 }.   Equivalently one may write
\bea
j_{1,2}={   P   \pm \sqrt{   P^2+1728 \, a^3\, d\, d' }\over 216\, d\, d'}     \label{sigmatauz}
\eea
 with $P=108 \,d d'-4 a^3-27\, b^2$. The functions $\tau_i(z)$ can be determined from the $j_i(z)$ given by
 (\ref{sigmatauz}) up to U-duality frame rotations (\ref{gamma22}).  It is important to notice that the functions
 $j_i(z)$ defined in (\ref{sigmatauz}) are not single valued. Indeed, going around  the zeros of the polynomial under the square
 root,  $j_1(z)$ and $j_2(z)$ get exchanged, i.e.  $\sigma \leftrightarrow \tau$.   This $\Z_2$ monodromy is however
 part of the U-duality group and therefore the functions assuming values in the quotient space (\ref{moduli22}) are single valued . The U-fold solution
 may be then viewed as a T-fold where 3- and 7-branes are identified.

Finally, one can determine the locations of the branes by looking for those points in the $z$-plane where one of the torus fibres degenerates,
let us say $j_1 \to \infty$. This happens at the $24$ zeros of the polynomial $d d'$.  At these points, writing $d d'\sim z-z_0$
one finds
\be
\tau_1(z)\approx  -{1\over 2\pi i } \log j_1(z) \approx {1\over 2 \pi } \log (z-z_0)  \, ,
 \ee
 so that $\tau\rightarrow \tau +1$ while the $\sigma$ modulus stays finite.
    We hence have $24$ 7-branes. We notice that an equivalent solution with only 3-branes can be written by flipping
    the signs in (\ref{sigmatauz}).
The two solutions  are clearly equivalent since 3- and 7-branes are identified.

 We remark that monodromies can be alternatively extracted by looking at the geometry of the K3 fibre near its degeneration points.
As explained in appendix \ref{appe8e8case}, the transcendental lattice of a K3 surface given by
\eqref{K3e8e8eq} is $U^{\oplus 2}$. Choosing an integral basis of cycles $\alpha_i$, we may write the holomorphic two-form as
\begin{equation}\label{omegaexpe8e8}
 \Omega= \tau \alpha_1 + \sigma \alpha_2 + \alpha_3- \tau\sigma\alpha_4 \, .
\end{equation}
In particular, this lattice contains a cycle $\alpha_1-\alpha_2$ which squares to $-2$, i.e. it can be represented by an $S^2$. Whenever $\tau=\sigma$, this
cycle collapses. As can be seen from \eqref{sigmatauz}, this happens precisely when $ P^2+1728 \, a^3\, d\, d'=0$. A collapsing $S^2$ gives
rise to the Picard-Lefschetz monodromy (see appendix for details)
 \begin{equation}
 \alpha_1\quad\leftrightarrow\quad \alpha_2 \, ,
\end{equation}
i.e. the roles of $\tau$ and $\sigma$ are exchanged. The same result is obtained upon collapsing $\alpha_3-\alpha_4$.
This argument shows that the presence of the T-duality transformation is linked to the existence of a cycle squaring to $-2$
in the transcendental lattice. Note that such a cycle is not necessarily present in any choice of transcendental lattice we can
make. In fact, the example presented in section \ref{fib}, where the transcendental lattice is $U(2)^{\oplus 2}$, does not allow
a Picard-Lefschetz monodromy. Correspondingly, there is no monodromy exchanging $7$ with $3$-branes there.

Summarizing, the $E_8\times E_8$ K3 fibration codifies a non-geometric  U-fold solution with non-trivial T-duality monodromies.
Consistently, the resulting string vacuum does not admit a weak coupling description in terms of D-branes as expected from
the underlying exceptional symmetry.

\subsection{$E_8\times E_7$ fibration: three moduli}

A deformation of the elliptic threefold presented in the last section is given by
 \bea
 f &=&  a_8(z) \, w^4 +c_{8+n}(z)\, w^3  \nn\\
 g &=&   d_{12- n}(z) w^7 + b_{12} (z)\, w^6+ d'_{12+ n}(z) w^5
 \eea
As before this threefold is given by a fibration of an elliptic K3 surface $X$ over $S^2$. This K3 surface is a one-parameter
deformation of the K3 surface with two singularities of type $E_8$ discussed in the last section, so that it has 3 complex structure moduli.
It has two singularities of the types $E_7$ and $E_8$ at $w=0$ and $w=\infty$, respectively.
The moduli space of complex structure deformations is given
by (\ref{moduli23}). We notice that this moduli space is isomorphic to the space of complex structures for a Riemann surface of genus
two\footnote{In fact, this $K3$ surface has a `Shioda-Inose structure'\cite{M1}, i.e. it is the double cover of a Kummer
surface which is constructed from a genus two curve \cite{AKumarg2}. } whose period matrix is
 \be
\Pi=\left(\begin{array}{cc} \tau & \beta \\
\beta & \sigma \end{array}\right)\, .
\ee
The U-duality group is then identified with the genus two modular group $Sp(4,\R)$. The precise map between $a,b,c,d,d'$ and the modular
forms of the genus two surface has been worked out in \cite{kumarelkies}
\be
a_8=-{\psi_4\over 3}  \qquad  b_{12} ={2\, \psi_6\over  27}  \qquad c_{8+n} \,d_{12-n}=2^{12}\,  \chi_{10}  \qquad d_{12-n}\, d'_{12+n}=2^{12} \, \chi_{12} \label{abcd0}
\ee
with $\psi_4,\psi_6$ the genus two Einstein series of weight $4,6$ and $\chi_{10}, \chi_{12} $ the cusp forms of weight $10$, $12$ (see the Appendix in \cite{Martucci:2012jk} for a self-contained review of genus two Riemann surfaces and modular forms).
We notice that the degrees of the $z$-polynomials in the left hand side of (\ref{abcd0}) are twice the weight of the
corresponding modular form in the right hand side as expected from the topological analysis in \cite{Martucci:2012jk}. The case $E_8\times E_8$ correspond
to the choice $c=0$ ( or $\beta=0$) where the genus two surface factorizes into a product
of two tori
 \bea
 \psi_4 &\to & E_4(\sigma) E_4(\tau)\qquad  \psi_6 \to  E_6(\sigma) E_6(\tau)\nn\\ \chi_{10} & \to & 0 \qquad ~~~~~~~~~~~~~~\chi_{12} \to \eta^{24} (\sigma) \eta^{24} (\tau)
 \eea
  and equations (\ref{abcd0}) reduce to (\ref{j1j2}). If one sets
  \be
  q_1=e^{2\pi i \tau} \qquad q_2=e^{2\pi i \sigma}   \qquad y=e^{2\pi i \beta}
  \ee
  and uses the expansions ( for small values of $q_1,q_2$) of  the cusp forms
  \be
 \chi_{10} = {(1-y)^2\over 4y} \,q_1 q_2+\dots\qquad
  \chi_{12} ={(1+10 y+y^2)\over 12y} \,q_1 q_2+\dots
 \ee
one finds now $q_1 q_2 y^{-1} \sim (z-z_0)$ near  the 24 zeros of $d d'$. Going around these points one finds
that the combination $\sigma+\tau-\beta$ is shifted by one, signalling for the presence of a brane of one of the three types. Locally, one can always choose a
frame so that the brane is a D7 or a D3 brane but globally the total charge should add to zero so exotic branes should be always present. Similarly, singularities
occur at the 20 zeros of $c d$ where $\chi_{10}$ vanishes. The structure of the singularity is more involved and a detailed description of the brane content for a general choice of the $a,b,c,d,d'$ polynomials is a challenging task that goes beyond the scope of this paper.

Still some information can be extracted  again from the geometry of the transcendental lattice $U^{\oplus 2}\oplus (-2)$.  We may write:
\begin{equation}
 \Omega= \tau \alpha_1 + \sigma \alpha_2 + \alpha_3 - (\tau\sigma-\beta^2) \alpha_4 +\alpha_5 \beta\, .
\end{equation}
Note that we have adjusted the coefficient of $\alpha_4$ to maintain $\int_{K3}\Omega\wedge\Omega=0$. In particular we find again that
there is a monodromy exchanging $\tau\leftrightarrow\sigma$ coming from the collapsing cycle $\alpha_1-\alpha_2$. Now, however, there
are further cycles of self-intersection $(-2)$ , e.g. $\alpha_5$, giving rise to a Picard-Lefschetz monodromy.  A systematic study of the richer
geometry of this fibration would be very welcome.

\subsection{$D_4^4$-fibration: two moduli }\label{fib}

\subsubsection{The $F_0$-case}

Next we consider an elliptic $K3$ surface given by a Weierstrass model at the sublocus of the moduli space  where
\bea
 f_{8,8}(w,z) &=& \alpha_{2k}(z) \, h_{4,4-k}(w,z)^2   \nn\\
  g_{12,12}(w,z) &=& \beta_{3k}(z)  \, h_{4,4-k}(w,z)^3  \, .  \label{fg}
\eea
The subscripts of the  various  polynomials here denote their degrees in the variables $w,z$.
 At each point $z$, the K3 fibre is singular in four points $w_i$, the zeros of $h$. According to Kodaira's classification
 of Table \ref{kodaira} they  correspond to four $D_4$ singularities.

  The holomorphic two-form (\ref{omega2}) factorizes into
\begin{align}
\Omega
& = \frac{d  x'}{\sqrt{x^{'3}+ \alpha x'  + \beta}} \quad\wedge \quad \frac{dw}{\sqrt{h}}\, ,
\end{align}
with $x \equiv x' h$.    On the other hand for $h(w,z)$ one may write
\begin{equation}
 h= a_4 w^4+4 a_3 w^3+6 a_2 w^2+4 a_1 w+ a_0     \, ,  \label{htorus}
\end{equation}
with the $a_i$'s being polynomials of degree $(4-k)$ in $z$.  Since $h$ is quartic in $w$,
one can identify
$\frac{dw}{\sqrt{h}}$ with the differential form of a torus with defining equation $\xi^2= h$, or in
 its  Weierstrass form
\be
\xi^2=w^3+ \tilde \alpha w+\tilde \beta \, ,
\ee
with
\bea
\tilde \alpha=4(4 a_1 a_3-a_4 a_0-3 a_2^2)   \qquad \tilde \beta=16(a_4 a_1^2+a_0 a_3^2+a_2^3 -a_4 a_2 a_0-2 a_1 a_2 a_3 ) \, .
\eea
  Notice that $\tilde \alpha$ and $\tilde \beta$ are polynomials of degree $(8-2k)$ and $(12-3k)$
  respectively.
 One can then identify $(\sigma,\tau)$ as the complex structures of  two tori. Explicitly
 \be
 j(\tau) = {4 (24 \alpha)^3 \over 4 \alpha^3+27 \beta^2 }     \qquad    j(\sigma) = {4 (24 \tilde\alpha)^3 \over4 \tilde \alpha^3+27 \tilde \beta^2} \label{jj} \, .
 \ee
The discriminants of the two tori are given by
 \be
  \Delta_\tau = 4 \alpha^3+27 \beta^2   \qquad    \Delta_\sigma= 4 \tilde \alpha^3+27 \tilde \beta^2 \, .
   \ee
  We notice that $ \Delta_\tau$ and  $\Delta_\sigma$ are polynomials of order $6k$ and $(24-6k)$ respectively with
zeros at the positions of the 3- and 7-branes\footnote{Here for simplicity we are referring as 3-branes to all exotic R-duals of $(p,q)$ 7-branes.}. We conclude then that there  are
  $24$ branes in total, $6k$ 7-branes and $24-6k$ 3-branes.
    Finally we observe that, when a 7- and a 3- branes collide, i.e. for  $\Delta_\tau=\Delta_\sigma=0$ the full K3 fibre
    degenerates\footnote{A K3 is singular if  $P=\partial_y P=\partial_x P=\partial_w P=0  $
with  $P=y^2-x^3-\alpha h^2 x-\beta $ the defining equation.
The vanishing of $P=\partial_y P=\partial_x P$  follows from $\Delta_\tau=0$ while
$\partial_w P\sim \partial_w h=0$
is  implied by $\Delta_\sigma=0$.  }.

  The transcendental lattice is now given by $U(2)^{\oplus 2}$ with intersection matrix $(^{02}_{20})$.
Choosing an integral basis $\alpha_i, i=1,\cdots,4$, we may write
\begin{equation}
 \Omega= \tau \alpha_1 + \sigma \alpha_2 + \alpha_3- \tau\sigma\alpha_4 \, .
\end{equation}
At first sight, this looks very similar to the situation with two $E_8$ singularities given by \eqref{omegaexpe8e8}. However,
the inner form on $T_X$ is now twice the one we had there and no  cycle
of self-intersection $-2$ is found. In particular, $\alpha_1-\alpha_2$ does not correspond to a sphere, but it is a reducible cycle
of self-intersection $-4$. This means that the K3 fibre is not singular when $\tau=\sigma$ and
therefore the solutions  with 3- and 7-branes remain distinct.

\subsubsection{The $F_n$-case}

 It is straightforward to generalize the previous analysis to the case when the base is an $F_n$ Hirzebruch surface with $n>0$.
One takes again
\bea
 f &=& \alpha (z) \, h(w,z)^2   \nn\\
  g  &=& \beta (z)  \, h(w,z)^3   \, , \label{fg_n}
\eea
but now with
\bea
\alpha(z) =\sum_{i=0}^{2k}   \alpha_i z^i    \qquad \beta=\sum_{i=0}^{3k}   \beta_i z^i \qquad
h(w,z) = \sum_{i=0}^{4}  \sum_{j=0}^{4-k+(2-i)n} h_{i,j}  w^i z^j \, .
\eea
 The only difference with the $F_0$ case is that now the coefficients $a_i$ entering in the expansion of $h(w)$ (\ref{htorus})
 are polynomials of degree $(4-k+(2-i)n)$ in $z$. Still the resulting $\tilde\alpha$ and $\tilde\beta$ are
 again given by polynomials of degree $(8-2k)$ and $(12-3k)$
  respectively and therefore the number of branes does not depend on $n$.

\subsubsection{Weak coupling limit }\label{fib2}

Finally we consider the special limit of the geometry where the axio-dilaton or four-form warping fields become almost
constant and large along the $z$-plane. In this limit one expects that the vacuum admits a perturbative description in terms
of D3, D7-branes and O3-, O7-planes.  Indeed if one sets $k=4$, the function $h$ becomes $z$-independent and therefore
the complex structure $\sigma$ of the $w$-torus is constant everywhere.  The threefold becomes
  $T^4/\Z_2\times T^2$  and describes the standard F-theory compactification on K3.
  The cases $k=0,1,2,3$ are new in the sense that some stacks of six 7-branes are commuted into stacks of 3-branes.
  In analogy with F-theory, one can ask whether the moduli of the Calabi-Yau can be tuned in such a way that $\sigma$ and $\tau$ are both constant and large almost everywhere in the $z$-plane allowing for a perturbative description in   type IIB theory.
 The answer is clearly yes, since we have two factorized tori, each of which can be treated in the same way  that the elliptic fibre
 of the more familiar $D_4$ F-theory geometry. For example, for the $\tau$ torus, the weak coupling limit corresponds to take
   \cite{Sen:1997gv}
   \bea
\alpha &=&  (-3 \gamma^2 +\epsilon\, \delta)   \nn\\
\beta &=&  (-2  \gamma^3+\epsilon \, \gamma \, \delta- \ft{1}{12}\epsilon^2 \chi )  \label{fg2} \, ,
\eea
  with $\epsilon$ a small constant and $\gamma,\delta,\chi$ two homogeneous functions of
degrees $k$, $2k$ and $3k$ respectively.  Plugging this into (\ref{jj}) one finds
\be
j(\tau)={(24)^4\over 2} {\gamma^4 \over \epsilon^2 (\delta^2- h \chi) } \, .
\ee
In the limit $\epsilon\to 0$, $\tau\to i \infty$ almost everywhere except at the
zeros of $\gamma$. Writing
\be
\gamma=\prod_{i=1}^k ( z-\zeta^O_i)   \qquad  (\delta^2- h \chi)=\prod_{m=1}^{4k} ( z-\zeta^D_m) \, ,
\ee
 we can identify then $\zeta^O_i$ as the position of the $k$ O7-planes and $\zeta^D_m$ as the positions
 of the $4k$ D7-branes. Indeed,  $j(\tau)$ has zeros of order four at $z=\zeta_i^O$ and therefore going around
 these points  $\tau \to \tau -4$ indicating the presence of an O7 plane at this point.
  Similarly one finds
  the holonomy $\tau\to \tau+1$ around $z=\zeta^D_m$ indicating the presence of a D7-brane.

   The analysis here can be repeated for the  $\sigma$-torus leading to identical conclusions. One finds $(4-k)$ groups of
   O3-planes (each group with charge -4) and $16-4k$ D3-branes.   Functions $\sigma(z)$, $\tau(z)$ codified in the geometry describe
   the exact running of the couplings in the 7-brane and 3-brane world-volume theories. Remarkably, like in the F-theory
   case, the full tower of multi-instanton corrections to these couplings can be extracted from this functional dependence.
\vspace{1cm}

\centerline{\large\bf Acknowledgements}

\vspace{0.5cm}
We thank L. Andrianopoli, M. Grana, M. Trigiante and  D. Waldram for discussions.
 This work is partially supported by the ERC Advanced Grant n.226455 ``Superfields" and by the Italian MIUR-PRIN contract 20075ATT78.
The work of A.P.B is supported by the STFC under grant ST/J002798/1.

\vspace{0.5cm}

\begin{appendix}

\section{Elliptic K3 surfaces}\label{app_ell_k3}
\label{sk3}

In this appendix, we collect some background material on elliptic K3 surfaces needed in the main text. We will not attempt
to provide a self-contained or through discussion and we will simply state the relevant facts. Further
details can be found in \cite{peters,schuettshioda,Aspinwall:1996mn} and references therein.

In this work, we restrict ourselves to elliptic K3 surfaces described by a Weierstrass equation. Such K3 surfaces
can be described as hypersurfaces in an ambient toric  space with the weight system
\begin{equation}
 \begin{array}{ccccc}
  y  & x & u & w_0 & w_1 \\
  \hline
  3 & 2 & 1 & 0 & 0 \\
  6 & 4 & 0 & 1 & 1
 \end{array}\, .
 \label{weights2}
\end{equation}
by an equation of the type
\begin{equation}\label{eqweierstrassK3app}
 y^2 = x^3 + x u^4 f_8  + u^6 g_{12} \, .
\end{equation}
Here, $f$ and $g$ are homogeneous functions of $(w_0,w_1)$ of the indicated degree.
From the homogeneity degree of the above equation and the weight system it follows that the hypersurface is a complex two-dimensional
Calabi-Yau manifold, i.e. a K3 surface.

The fact that we have an elliptic fibration can also be instantly verified:
if we fix any $[w_0:w_1]$, \eqref{eqweierstrassK3app} describes an elliptic curve, i.e. a torus. Hence the elliptic
K3 surface is given by a torus sitting over every point of an $S^2$. Furthermore, there is a (holomorphic) section:
we may embed the base $S^2$ of the elliptic fibration by mapping it to the point $[x:y:u]=[1,1,0]$ in the fibre.

As it is true for any Calabi-Yau manifold, a K3 surface has a non-vanishing holomorphic top-form $\Omega$.
By the method of residues (see e.g. \cite{GH}), the holomorphic two-form $\Omega$ of a K3 surface embedded in an ambient
toric variety map be written as (see \cite{Denef:2008wq} for a nice derivation)
 \be
 \Omega={1\over 2 \pi i} \oint_{P=0} {w \over P} \cdot \prod_a V_a
 \ee
 with $w/P$ an invariant top form in the ambient space
 \be
w  = dx \,dy\, du \, dw_0\, dw_1     \qquad ~~~~~~~~~~~~P= y^2-(x^3+f x+g)
\ee
and
\bea
V_1 &=& 3 \, y\, \partial_y +2 \, x\, \partial_x +u\, \partial_u \nn\\
V_2 &=& 6 \, y\, \partial_y +4 \, x\, \partial_x +w_0\, \partial_{w_0}+w_1\, \partial_{w_1}
 \eea
 the generators of the $C^*$'s actions in (\ref{weights2}).
  In the chart where $u=w_0=1$, $w_1=w$, one can then write
  \bea
 \Omega={1\over 2 \pi i} \oint_{P=0} { dx \,dy\, dw\over y^2-  (x^3+f x+g)} ={ dx \, dw\over \sqrt{x^3+f x+g} }
 \eea
 The fibre of the elliptic fibration is smooth for a generic point of the base $S^2$. Over the $24$ points where the discriminant
\begin{equation}
\Delta=4f_8^3+27g_{12}^2
\end{equation}
vanishes, however, the fibre degenerates by pinching the one cycle $(p,q)$. If we encircle one of those points, the fibre undergoes
the $SL(2,\Z)$ monodromy transformation  $(^{1-pq~p^2}_{
	 -q^2~1+pq})$\footnote{In F-Theory, such degenerations give the locations of $(p,q)$ 7-branes.}.
 As long as these degenerate fibres stay separate, the K3 surface stays smooth. When the polynomials $f_8$ and $g_{12}$
are such that two or more of these singular fibres go on top of each other, i.e. $\Delta$ has a double (or higher) root,
also the K3 surface becomes singular. The singularities that occur in this way are nothing but the ADE (or simple surface, Kleinian,
duVal) singularities. This is not unexpected, as these are precisely the orbifold singularities for which the orbifold group
is a finite subgroup of $SU(2)$, the holonomy group of a K3 surface.
The types of singular fibres that can occur and the corresponding
ADE singularities are displayed in Table \ref{kodaira}.

 \begin{table}
 \centering
\begin{tabular}[h]{|c|c|c|c|c|c|c|}
\hline
 $\mbox{ord}(f)$ & $\mbox{ord}(g)$ & $\mbox{ord}(\Delta)$ & fibre type & singularity type & monodromy \\ \hline \hline
 $\geq 0$ & $\geq 0$ & $0$ & smooth & none &  $\left(\begin{array}{cc} 1 & 0 \\ 0 & 1 \end{array}\right)$ \\ \hline
$0$&$0$&$n$&$I_n$&$A_{n-1}$ & $\left(\begin{array}{cc} 1 & n \\ 0 & 1 \end{array}\right)$   \\ \hline
$2$ & $\geq 3$ & $n+6$ &  $I_n ^*$ & $D_{n+4}$ & $-\left(\begin{array}{cc} 1 & n \\ 0 & 1 \end{array}\right)$ \\ \hline
$\geq 2$ & $3$ & $n+6$ &  $I_n ^*$ & $D_{n+4}$ & $-\left(\begin{array}{cc} 1 & n \\ 0 & 1 \end{array}\right)$\\ \hline
$\geq 1$ & $1$ & $2$& $II$ & none &$\left(\begin{array}{cc} 1 & 1 \\ -1 & 0 \end{array}\right)$\\ \hline
$\geq 4$ & $5$ & $10$ & $II^*$  & $E_8$ & $\left(\begin{array}{cc} 0 & -1 \\ 1 & 1 \end{array}\right)$\\ \hline
$1$ & $\geq 2$ & $3$ &  $III$ & $A_1$ &$\left(\begin{array}{cc} 0 & 1 \\ -1 & 0 \end{array}\right)$\\ \hline
$3$ & $\geq 5$ & $9$ &  $III^*$  & $E_7$ &$\left(\begin{array}{cc} 0 & -1 \\ 1 & 0 \end{array}\right)$\\ \hline
$\geq 2$ & $2$ & $4$ &   $IV$  & $A_2$ &$\left(\begin{array}{cc} 0 & 1 \\ -1 & -1 \end{array}\right)$\\ \hline
$\geq 3$ & $4$ & $8$ &  $IV^*$ & $E_6$&$\left(\begin{array}{cc} -1 & -1 \\ 1 & 0 \end{array}\right)$\\ \hline
\end{tabular}
\caption{Kodaira's classification of bad fibres in terms of the vanishing degree of $f$, $g$ and $\Delta$. Also given is the corresponding monodromy and the type of surface singularity. This table has already appeared in \cite{Morrison:1996pp}.}
\label{kodaira}
\end{table}

The Euler characteristic of a K3 surface is $24$. Besides the 0- and 4-form there
   are  $20$ harmonic forms of type $(1,1)$ as well as the holomorphic $(2,0)$ form and its complex conjugate.
In the following, however, we will be mostly interested in the integer (co)homology $H^2(K3,\mathbb{Z})$. One might think
of the integer homology as being generated by (homology classes of) submanifolds of the K3 surface via the natural pairing
between homology and cohomology given by the integration. The integer cohomology classes $H_2(K3,\mathbb{Z})$ are those ones for which
the integral over any submanifold gives an integer number, i.e. it is the dual lattice. By Poincar\'e duality, the two lattices are
isomorphic. This implies we have an inner form $\alpha\cdot\beta$ on both of them. Thinking in terms of homology, this number counts
the geometric intersections between two representatives, in terms of cohomology this number is found by wedging the two forms and
integrating over the whole K3 surface:
\begin{equation}
 \alpha\cdot\beta = \alpha\cap\beta=\int_{K3}\alpha\wedge\beta \, .
\end{equation}
Here, we both denote an element of $H_2(K3,\mathbb{Z})$ and its Poincar\'{e} dual in $H^2(K3,\mathbb{Z})$ by the same letter.
With this inner form,
\begin{equation}\label{h2k3zapp}
H^2(K3,\mathbb{Z})=U^{\oplus 3}\oplus E_8^{\oplus 2} \, ,
\end{equation}
where $U$ is the lattice with inner form $(^{01}_{10})$ and $E_8$ is the root lattice of $E_8$\footnote{More precisely we denote by $E_8$ the lattice with   intersection matrix given by minus the Cartan matrix of the $E_8$ Lie algebra.}.
We will refer to elements of this lattice as cycles. Given an element $\gamma$ in $H_2(K3,\mathbb{Z})$ which can be represented by a
Riemann surface of genus $g$, the self-intersection number is simply
\begin{equation}
 \gamma\cdot\gamma = 2g-2 \, .
\end{equation}
Hence a sphere will correspond to a lattice point with self-intersection $-2$ and a torus will have self-intersection zero.
Notice that this means that a cycle of self-intersection smaller than $-2$ can never correspond to an irreducible submanifold.

The Picard lattice of a K3 surface $X$ is defined as
\begin{equation}
 \mbox{Pic}(X)=H^2(X,\mathbb{Z})\cap H^{1,1}(X) \, ,
\end{equation}
so that it contains only integral cycles of type $(1,1)$. Its dimension is called the Picard number $\rho$.
Clearly, any two-cycle given by an algebraic equation becomes a member
of the Picard lattice. By the Lefschetz theorem on $(1,1)$ classes, this statement also has a converse: the Picard lattice is
generated by algebraic cycles.

The transcendental lattice is defined as the orthogonal complement of the Picard lattice in $H^2(X,\mathbb{Z})$:
\begin{equation}\label{defTX}
 T_X=\mbox{Pic}^\perp \subset H^2(X,\mathbb{Z})\, .
\end{equation}

As it is familiar from Calabi-Yau threefolds, the moduli space of a K3 surface defined in terms
of algebraic equations consists of K\"ahler and complex structure deformations. Using the Picard and the transcendental lattices we may
write the K\"ahler from $J$ and the holomorphic two-form $\Omega$ as
\begin{align}\label{jwexpapp}
J&=\sum_{\delta_i\in\mbox{Pic}(X)}j_i\delta_i \quad\quad j_i\in \mathbb{R} \\
\Omega&=\sum_{\gamma_i\in T_X}w_i\gamma_i \quad\quad w_i\in \mathbb{C} \, ,
\end{align}
where
\begin{equation}
\int_{X} \Omega\wedge\Omega = 0
\end{equation}
puts an extra constraint on the $w_i$. The mutual orthogonality of the Picard lattice and the transcendental lattice
ensures that $J\cdot\Omega=0$, as it should be. From \eqref{jwexpapp}, the periods
\begin{equation}
\pi_i = \int_{C_i\in H_2(X,\mathbb{Z})}\Omega \, ,
\end{equation}
are determined by using \eqref{h2k3zapp}. Fixing the Picard lattice, the moduli space of complex structures becomes the Grassmannian
\begin{equation}\label{modalgk3app}
{\rm O}(T_X)\backslash { {\rm O}(2,20-\rho;\mathbb{R})\over
 {\rm O}(2;\mathbb{R}) \times {\rm O}(20-\rho;\mathbb{R})}\, ,
\end{equation}
where $\rho$ denotes the Picard number and ${\rm O}(T_X)$ denotes the isometries of the transcendental lattice.

A K3 surface develops a singularity of type ADE when there is two cycle $\alpha$ with $\alpha^2=-2$ for which
\begin{equation}
J\cdot\alpha=\Omega\cdot\alpha=0 \, .
\end{equation}
Intuitively, this means that $\alpha$, which correspond to an $S^2$, has collapsed to a point. For this reason, cycles of
this type are referred to as vanishing cycles. The vanishing cycles generate a root lattice and the lattices obtained
in this way precisely match the ADE type of the corresponding singularity. If the lattice of vanishing cycles decomposes
into a direct sum of root lattices, the corresponding K3 surface has distinct singularities of the corresponding types.

We have discussed the Picard and the transcendental lattices for smooth K3 surfaces above. In the singular case,
it is customary in the mathematics literature (and natural from several viewpoints) to include the vanishing cycles in
the Picard lattice. We can rephrase this in the following way: for any ADE singularity, there is a unique resolution
which corresponds to a K\"ahler deformation. Hence we can use the corresponding resolved K3 surface instead of the singular one to
define the Picard and transcendental lattices. But this simply means grouping the vanishing cycles with the Picard lattice.

When we consider fibrations of K3 surfaces, over an $S^2$ parametrized by $[z_0:z_1]$, say, the complex
structure of the K3 fibre varies from point to point. We may think of this either as making the coefficients appearing
in $f_8$ and $g_{12}$ functions of $[z_0:z_1]$ or as making the periods, i.e. the coefficients $w_i$ in (\ref{jwexpapp}), functions of $[z_0,z_1]$.
From the latter perspective, it is clear that there will  be (in general) points for which the K3
surface develops a singularity of ADE type. Encircling such a locus, there will be a monodromy transformation
acting on $H^2(K3,\mathbb{Z})$. Since the vanishing cycles at these points is a sphere, the monodromy action can be determined
from the intersection form through the
Picard-Lefschetz formula (see e.g. \cite{arnold}). If a cycle $\gamma$ shrinks at the point we are encircling, the
induced map on any other cycle $\alpha$ is given by
\begin{equation}\label{piclefapp2}
\alpha\mapsto \alpha + (\alpha\cdot\gamma)\gamma \, .
\end{equation}
Note that, for root lattices, this is nothing but a Weyl reflection, which is a symmetry of the root lattice.
This fits with the structure of the moduli space \eqref{modalgk3app}, in which the isometries of the transcendental
lattice are modded out. Note however, that not all the isometries of the transcendental lattice are Weyl reflections
of the transcendental lattice and the vanishing of cycles with topology different from the sphere can (and indeed) occur.
In the text, we mainly determine the monodromy action from the explicit expression for the periods.

\subsection{Picard and transcendental lattices of generic elliptic K3 surfaces}

For an elliptic K3 surface (with $f_8$ and $g_{12}$ chosen generically), the Picard lattice is generated by the base and the fibre
of the elliptic fibration.
Let us denote the divisor classes associated to the basic hyperplanes $x_i=0$ by $D_{x_i}$. The invariance under
$\mathbb{C}^{*2}$ of $y^2/x^3$, $y^2/(u^6 w^{12})$, and $w_0/w_1$ imply the relations
\be
D_{w_0}=D_{w_1}=D_w  \qquad 2D_y=3 D_x=6 D_u+12 D_w   \label{rel}
\ee
To find the intersection numbers is convenient to think of the ambient space $W$ as the $U(1)^2$
quotient of the hyperplane defined by the $U(1)^2$ moment maps
\bea
&& 3| y|^2+2 |x|^2+|u|^2=\xi_1 \nn\\
 && 6| y|^2+4 |x|^2+|w_0|^2+|w_1|^2=\xi_1
\eea
 Then it is easy to see that equations $x=y=u=0$ and $w_1=w_2=0$ have no solution while $x=y=w=0$ has a single solution.
 This implies
 \be
 \int_W D_w^3=  \int_W D_w^2 \,D_u= \int_W D_x \,D_y \, D_u=0  \qquad  \int_W D_x \,D_y \,D_w=1 \label{w3}
 \ee
 with $W$ the threefold.
   Combining (\ref{rel}) and (\ref{w3}) one finds
 \be
 \int_W D_w^3=  \int_W D_w^2 \,D_u= 0  \qquad \int_W D_u^2 \,D_w =\ft16   \qquad  \int_W D_u^3 =-\ft23 \label{w32}
 \ee
 The intersection on $K3$ can be found from the adjunction formula
 \be
 \int_{K3} \alpha=\int_W \alpha \wedge (6 D_u+12 D_w)
 \ee
 leading to
  \be
 \int_{K3}  D_w^2=0  \qquad \int_{K3} D_u^2=-2  \qquad   \int_{K3} D_u\, D_w=1  \label{intk3}\, .
 \ee
This is as expected: \eqref{eqweierstrassK3app} defines a fibration of the torus $w=0$ over the sphere $u=0$,
which is also a section of the fibration, i.e. it intersects every fibre in a single point. If we choose a
different integral basis for the Picard lattice,
\begin{equation}
 \alpha_1= D_w\, , \quad\quad \alpha_2= D_u+D_w \, ,
\end{equation}
its inner form becomes $(^{01}_{10})$. From this is follows that the Picard lattice is $U$ and the transcendental lattice of
the generic elliptic K3 surface is $T_X=U^{\oplus 2}\oplus E_8^{\oplus 2}$ (the orthogonal complement of $U$
inside (\ref{h2k3zapp}) ).

\subsection{Picard and transcendental lattices of our examples}

In the following we describe in details the transcendental lattices for the three examples we discuss in this paper. We can restrict
the complex structure moduli space \eqref{modalgk3app} by enlarging the Picard number $\rho$. As the examples we
are considering have 2 or 3 complex structure moduli, the Picard numbers are 18 or 17.
This can be achieved by considering elliptic K3 surfaces given by the blow up of an ADE singularity of high rank ($16$ or $15$).
As the blowups of the singularities we consider are unique and well known, we do not have to carry them out explicitly.
We will now consider our set of examples in turn. From what we have said, it is natural
to label the examples by their singularity structure.

\subsubsection{The $E_8  \oplus E_8$ case}\label{appe8e8case}

\begin{figure}[!h]
\begin{center}
\scalebox{.5}{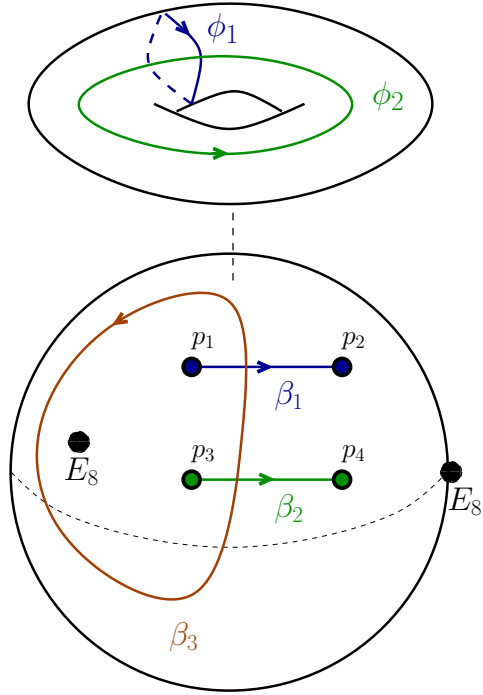}
\caption{A cartoon depicting an elliptic K3 surface with two singularities of type $E_8$.
\label{e8e8figure}}
\end{center}
\end{figure}

Let us consider the elliptic K3 surface
 \bea\label{eqe8e8app}
y^2=x^3+x \, a \, w^4 +( d w^7 + b\, w^6+d' w^5)
 \eea
with discriminant
\be
\Delta=w^{10} P_4(w) \, .
\ee
Here,
\begin{equation}
 P_4 = 27d'^2+54d'bw +(4a^3+27b^2+54d'd)w^2 +54 dbw^3 + 27d^2w^4  \, ,
\end{equation}
is a polynomial of order $4$ in $w$. Using the classification of Table \ref{kodaira} one finds two $E_8$ singularities over
$w=0$ and $w=\infty$ and four $A_0$ regular points associated to the zeros of $P_4(w)$, see Figure \ref{e8e8figure}.

As we have seen already, the sublattice of the Picard lattice generated by the fibre and section of the elliptic
fibration is $U$. Together with the vanishing cycles of the two $E_8$ singularities we hence find that \eqref{eqe8e8app} has the
Picard lattice
\begin{equation}
\mbox{Pic}=U\oplus  E_8 \oplus E_8 \label{pice82}\, .
\end{equation}

The transcendental lattice is $U^{\oplus 2}$,i.e.  the orthogonal complement of the Picard lattice (\ref{pice82}) in
$ U^{\oplus 3}\oplus E_8^{\oplus 2}$.
Following \cite{Aspinwall:1997eh}, see also \cite{Gaberdiel:1997ud,DeWolfe:1998pr,Hayashi:2010zp}, we may construct
this lattice as follows. Let us denote the four roots of $P_4$ by $p_i$,
$i=1..4$. The one-cycles in the $T^2$ fibre which collapse over these four points are pairwise the same,
for $p_1$ and $p_3$ a cycle $\phi_1$ shrinks and for $p_2$ and $p_4$ a one-cycle $\phi_2$ shrinks. We may then choose a basis
such that $\phi_1$ and $\phi_2$ are as depicted in Figure \ref{e8e8figure}. Hence we may construct a two-cycle $\gamma_1$ by fibring
$\phi_1$ over the interval $\beta_1$ connecting $p_1$ and $p_3$. This cycle has
the topology of a two-sphere and therefore its self-intersection number is $-2$. Similarly, a second sphere is made of the fibration of
$\phi_2$ over the path $\beta_2$ connecting $p_2$ and $p_4$. Hence we may suggestively write
\begin{align}
 \gamma_1 = \beta_1\wedge\phi_1 \quad\quad \gamma_2 = \beta_2\wedge\phi_2 \, .
\end{align}

Furthermore, the $SL(2,\mathbb{Z})$ monodromies of this configuration are such that the monodromy map induced along
the loop $\beta_3$ is trivial\footnote{
This can be seen from the $\Z_2$ symmetry which exchanges the internal and external regions of the sphere surround by $\beta_3$. It
implies that the monodromy around $\beta_3$ should coincide with its inverse.}. Hence we can fibre any one-cycle in the elliptic fibre over $\beta_3$ to
obtain a two-cycle. Using the same basis of cycles in the fibre as before, we can hence form the two cycles
\begin{align}
 \alpha_1 = \beta_3\wedge\phi_2 \quad\quad\alpha_3=\beta_3\wedge \phi_1\, ,
\end{align}
so that
\begin{equation}
 \alpha_1\cdot\gamma_1=1 \quad \alpha_3\cdot\gamma_2=1 \, .
\end{equation}
As the fibration of the elliptic fibre along $\beta_3$ is trivial, these cycles are just $S^1\times S^1$, i.e. they are
two-tori, so that
\begin{equation}
 \alpha_1^2=\alpha_3^2=0
\end{equation}
Finally, we may define
$\alpha_2=\gamma_1+\alpha_1$, $ \alpha_4=\gamma_2+\alpha_3$
so that the four two-cycles $\alpha_i$ have the intersection form
\begin{equation}
 T_X= \left(\begin{array}{cc}
  0 & 1\\
  1 & 0
 \end{array}\right) \oplus
 \left(\begin{array}{cc}
  0 & 1\\
  1 & 0
 \end{array}\right) = U^{\oplus 2}\, .
 \label{T_X_e8e8}
\end{equation}

For special values of $a,b,d,d'$, some of the roots of $P_4$ may come together, so that the $K3$ surface develops further
singularities. For example when $d'$ goes to zero, $p_1$ and $p_3$  will go to zero in the $w$-plane and so the
cycle $\beta_3$ is pinched. The same is true for $d$ going to zero where $p_2$ and $p_4$ go to infinity.
On the other hand when $a=0$, $P_4$ becomes a square, i.e. the $p_i$ will move pairwise on top of each other.
When $p_1$ and $p_2$ or $p_3$ and $p_4$ coincide,
 $\gamma_1$ or $\gamma_2$ collapse.
For instance when $\gamma_1=\alpha_2-\alpha_1$ collapses one finds the Picard-Lefschetz monodromy
\begin{equation}\label{piclefapp}
\alpha\mapsto \alpha + (\alpha\cdot\gamma_1)\gamma_1 \, .
\end{equation}
 inducing the map
\begin{equation}
 \alpha_1\quad\leftrightarrow\quad \alpha_2 \, .
\end{equation}

\subsubsection{The $E_8  \oplus E_7$ case}\label{appe8e7case}

\begin{figure}[!h]
\begin{center}
\scalebox{.5}{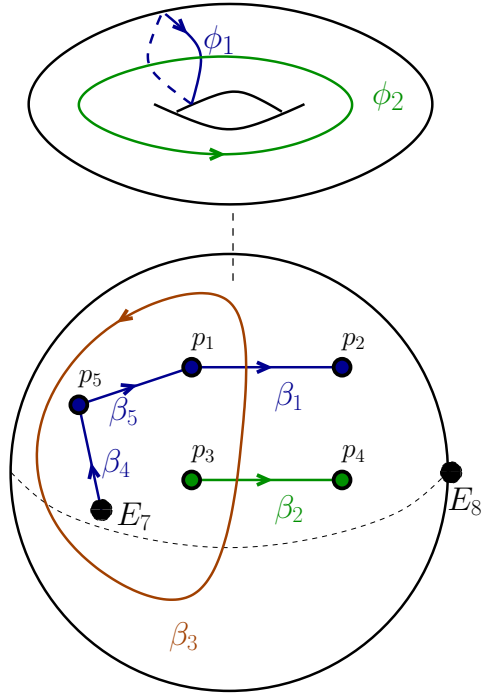}
\caption{ K3 surface with  one $E_8$, one $E_7$ singularity and five $A_0$ regular points.
\label{e8e7figure}}
\end{center}
\end{figure}

Next, we consider the elliptic K3 surface
 \bea\label{eqe7e8app}
y^2=x^3+x (\, a \, w^4+c \, w^3) +( d w^7 + b\, w^6+d' w^5)
 \eea
with discriminant
\be
\Delta=w^{9} P_5(w)\, ,
\ee
where
\begin{equation}\label{discre7e8app}
 P_5= -4c^3-12ac^2w-12a^2cw^2-w\left(27d'^2+54d'w(b+dw)+w^2(4a^3+27(b+dw)^2)\right) \,
\end{equation}
is a polynomial of order $5$ in $w$. According to Table (\ref{kodaira}) one finds one $E_8$ singularity at
$w=\infty$, one $E_7$ singularity at $w=0$ and five regular points at the zeros of $P_5(w)$. We have depicted
the situation in figure \ref{e8e7figure}.

The Picard lattice now becomes
 \begin{equation}
\mbox{Pic}=U\oplus  E_8 \oplus E_7 \label{pice81}
\end{equation}
so that the transcendental lattice must be five dimensional.

This configuration may be obtained from the situation with two $E_8$ singularities by a one-parameter deformation. The four cycles $\alpha_i$
constructed in the last section remain present and there is a single new cycle $\alpha_5$ in the transcendental lattice. From the explicit
construction of the vanishing cycles of an $E_8$ singularity in an elliptic surface \cite{Gaberdiel:1997ud}, one can infer that the cycle which is deformed
to finite volume simply connects $p_5$ to the remaining $E_7$ singularity, i.e. it is a $S^2$ which is dual to
$\beta_4\wedge\phi_1$.
Even though this cycle measures the deformation of an $E_8$ singularity to an $E_7$, it is not part of the transcendental lattice since it
intersects with the vanishing cycles of the $E_8$ singularity which are part of the Picard lattice.

On the other hand, $\gamma=\beta_5\wedge\phi_1$ does not intersect any cycle of the Picard lattice, so that it is contained in the transcendental lattice.
It has the following non-vanishing intersection numbers
\begin{equation}
 \gamma^2=-2 \, ,\quad \gamma\cdot\alpha_2=1 \, .
\end{equation}
We can choose $\alpha_i$, $i=1,2,3,4$ and $\alpha_5 \equiv \gamma - \alpha_1$ as the generators of the transcendental lattice. In this basis its inner form is
\begin{equation}
 T_X= U^{\oplus 2}\oplus (-2)\, .
 \label{T_X_e8e7}
\end{equation}

The K3 surface  \eqref{eqe7e8app} has in this case a very rich pattern of degenerations. Whenever $c=0$, we get back to the case with $E_8\times E_8$ singularities, i.e. the cycle $\gamma$ collapses. The remaining singular point can be  identified
by studying the discriminant of \eqref{discre7e8app}. A systematic study is beyond the scope of this paper.

\subsubsection{The $D_4^4$ case}

\begin{figure}[!h]
\begin{center}
\scalebox{1}{ 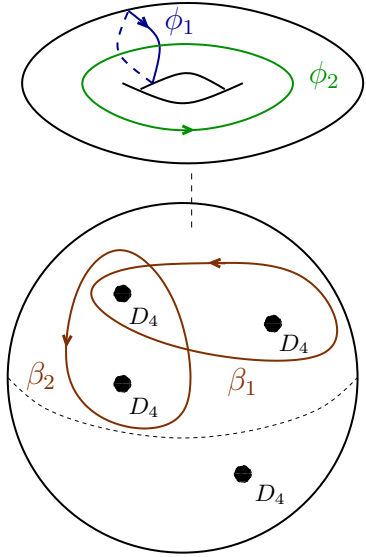 }
\caption{Elliptic K3 surfaces with four $D_4$ singularities. The basic two cycles are built by taking an arbitrary one-cycle in the fibre $\phi_{1,2}$ and transporting it along $\beta_1$ or $\beta_2$.}
\label{4d4figure}
\end{center}
\end{figure}

A situation with four singularities of type $D_4$ is given by
\be
y^2=x^3+x\, \alpha\, h^2  +\beta h^3
\ee
with $h(w)$ a polynomial of order 4 in $w$. The discriminant is
\begin{equation}
 h^6(4\alpha^2+27\beta^2)
\end{equation}
Using again Table (\ref{kodaira}), one can see that the K3 surface
has singularities of type $D_4$ at the four roots of $h(w)$ . Hence the Picard lattice contains\footnote{It turns out that this is not all of the Picard lattice, which also contains a few integral
cycles which are linear combinations of the elements of the lattices $U$ and $D_4^4$. As this subtlety is irrelevant for our analysis, we
do not dwell on this further.}
\begin{equation}
 U  \oplus D_4^{\oplus 4}   \label{picd44} \, .
\end{equation}
The transcendental lattice is then four-dimensional and can be constructed as follows  (see \cite{Braun:2008ua} for a similar
discussion). As the monodromy induced around the four $D_4$-singular points  fibres is $-1$, any path in the base
which encircles two such points will give trivial monodromy. A basis of such paths are, let us say $\beta_1$ and $\beta_2$, is depicted in Figure \ref{4d4figure}.
 A basis of two-cycles  is built by taking an arbitrary one-cycle in the fibre $\phi_{1,2}$ and transporting it along $\beta_1$ or $\beta_2$. Using dual forms instead of cycles (which be denote by the same letters) we can write
 \be
 \alpha_{ij}= \beta_i\wedge \phi_j   \qquad ~~~~~~~~i,j=1,2
\ee
 We notice that the cycles associated to $\alpha_{ij}$ have self-intersection zero since  two homologous cycles differing
 by the choice of a different  representative for $\phi_1$ or $\phi_2$ do not intersect. On the other hand
 $\alpha_{12}\cdot\alpha_{21} = \alpha_{11}\cdot\alpha_{22}=2$. The associated cycles intersect twice as can be seen in
 Figure \ref{4d4figure}  ($\beta_1$ and $\beta_2$ meet in two points with opposite orientation but the orientations of
 $\phi_1$ and $\phi_2$ are also flipped).
 Hence the inner product on the transcendental lattice takes the form
\begin{equation}
 T_X= \left(\begin{array}{cc}
  0 & 2\\
  2 & 0
 \end{array}\right) \oplus
 \left(\begin{array}{cc}
  0 & 2\\
  2 & 0
 \end{array}\right)
\end{equation}
The transcendental lattice is then $U(2)^{\oplus 2}$. As discussed e.g. in \cite{Braun:2008ua}, the same result for the transcendental lattice is obtained
by embedding the Picard lattice (\ref{picd44}) into $H^2(K3,\Z)=U^{\oplus 3}\oplus E_8^{2}$ and computing its orthogonal complement.

\end{appendix}
\providecommand{\href}[2]{#2}\begingroup\raggedright\endgroup


\begin{thebibliography}{10}

\bibitem{Vafa:1996xn}
C.~Vafa, \emph{{Evidence for F theory}},
  \href{http://dx.doi.org/10.1016/0550-3213(96)00172-1}{Nucl.Phys. {\bf B469}
  (1996)  403--418},
\href{http://arxiv.org/abs/hep-th/9602022}{{\tt arXiv:hep-th/9602022
  [hep-th]}}.
%%CITATION = HEP-TH/9602022;%%.

\bibitem{Sen:1996vd}
A.~Sen, \emph{{F theory and orientifolds}},
  \href{http://dx.doi.org/10.1016/0550-3213(96)00347-1}{Nucl.Phys. {\bf B475}
  (1996)  562--578},
\href{http://arxiv.org/abs/hep-th/9605150}{{\tt arXiv:hep-th/9605150
  [hep-th]}}.
%%CITATION = HEP-TH/9605150;%%.

\bibitem{Beasley:2008dc}
C.~Beasley, J.~J. Heckman, and C.~Vafa, \emph{{GUTs and Exceptional Branes in
  F-theory - I}}, \href{http://dx.doi.org/10.1088/1126-6708/2009/01/058}{JHEP
  {\bf 0901} (2009)  058},
\href{http://arxiv.org/abs/0802.3391}{{\tt arXiv:0802.3391 [hep-th]}}.
%%CITATION = ARXIV:0802.3391;%%.

\bibitem{Beasley:2008kw}
C.~Beasley, J.~J. Heckman, and C.~Vafa, \emph{{GUTs and Exceptional Branes in
  F-theory - II: Experimental Predictions}},
  \href{http://dx.doi.org/10.1088/1126-6708/2009/01/059}{JHEP {\bf 0901} (2009)
   059},
\href{http://arxiv.org/abs/0806.0102}{{\tt arXiv:0806.0102 [hep-th]}}.
%%CITATION = ARXIV:0806.0102;%%.

\bibitem{Donagi:2008ca}
R.~Donagi and M.~Wijnholt, \emph{{Model Building with F-Theory}},
  Adv.Theor.Math.Phys. {\bf 15} (2011)  1237--1318,
\href{http://arxiv.org/abs/0802.2969}{{\tt arXiv:0802.2969 [hep-th]}}.
%%CITATION = ARXIV:0802.2969;%%.

\bibitem{Hayashi:2008ba}
H.~Hayashi, R.~Tatar, Y.~Toda, T.~Watari, and M.~Yamazaki, \emph{{New Aspects
  of Heterotic--F Theory Duality}},
  \href{http://dx.doi.org/10.1016/j.nuclphysb.2008.07.031}{Nucl.Phys. {\bf
  B806} (2009)  224--299},
\href{http://arxiv.org/abs/0805.1057}{{\tt arXiv:0805.1057 [hep-th]}}.
%%CITATION = ARXIV:0805.1057;%%.

\bibitem{Billo:2011uc}
M.~Billo, M.~Frau, L.~Giacone, and A.~Lerda, \emph{{Holographic
  non-perturbative corrections to gauge couplings}},
  \href{http://dx.doi.org/10.1007/JHEP08(2011)007}{JHEP {\bf 1108} (2011)
  007},
\href{http://arxiv.org/abs/1105.1869}{{\tt arXiv:1105.1869 [hep-th]}}.
%%CITATION = ARXIV:1105.1869;%%.

\bibitem{Fucito:2011kb}
F.~Fucito, J.~Morales, and D.~R. Pacifici, \emph{{Multi instanton tests of
  holography}}, \href{http://dx.doi.org/10.1007/JHEP09(2011)120}{JHEP {\bf
  1109} (2011)  120},
\href{http://arxiv.org/abs/1106.3526}{{\tt arXiv:1106.3526 [hep-th]}}.
%%CITATION = ARXIV:1106.3526;%%.

\bibitem{Billo:2012st}
M.~Billo, M.~Frau, F.~Fucito, L.~Giacone, A.~Lerda, J.~Morales, and
  D.~Ricci~Pacifici, \emph{{Non-perturbative gauge/gravity correspondence in
  N=2 theories}}, \href{http://dx.doi.org/10.1007/JHEP08(2012)166}{JHEP {\bf
  1208} (2012)  166},
\href{http://arxiv.org/abs/1206.3914}{{\tt arXiv:1206.3914 [hep-th]}}.
%%CITATION = ARXIV:1206.3914;%%.

\bibitem{Martucci:2012jk}
L.~Martucci, J.~F. Morales, and D.~R. Pacifici, \emph{{Branes, U-folds and
  hyperelliptic fibrations}},
  \href{http://dx.doi.org/10.1007/JHEP01(2013)145}{JHEP {\bf 1301} (2013)
  145},
\href{http://arxiv.org/abs/1207.6120}{{\tt arXiv:1207.6120 [hep-th]}}.
%%CITATION = ARXIV:1207.6120;%%.

\bibitem{deBoer:2012ma}
J.~de~Boer and M.~Shigemori, \emph{{Exotic Branes in String Theory}},
\href{http://arxiv.org/abs/1209.6056}{{\tt arXiv:1209.6056 [hep-th]}}.
%%CITATION = ARXIV:1209.6056;%%.

\bibitem{Greene:1989ya}
B.~R. Greene, A.~D. Shapere, C.~Vafa, and S.-T. Yau, \emph{{Stringy Cosmic
  Strings and Noncompact Calabi-Yau Manifolds}},
\href{http://dx.doi.org/10.1016/0550-3213(90)90248-C}{Nucl.Phys. {\bf B337}
  (1990)  1}.
%%CITATION = NUPHA,B337,1;%%.

\bibitem{Becker:1996gj}
K.~Becker and M.~Becker, \emph{{M theory on eight manifolds}},
  \href{http://dx.doi.org/10.1016/0550-3213(96)00367-7}{Nucl.Phys. {\bf B477}
  (1996)  155--167},
\href{http://arxiv.org/abs/hep-th/9605053}{{\tt arXiv:hep-th/9605053
  [hep-th]}}.
%%CITATION = HEP-TH/9605053;%%.

\bibitem{Dasgupta:1999ss}
K.~Dasgupta, G.~Rajesh, and S.~Sethi, \emph{{M theory, orientifolds and G -
  flux}}, JHEP {\bf 9908} (1999)  023,
\href{http://arxiv.org/abs/hep-th/9908088}{{\tt arXiv:hep-th/9908088
  [hep-th]}}.
%%CITATION = HEP-TH/9908088;%%.

\bibitem{Bianchi:1999uq}
M.~Bianchi, J.~F. Morales, and G.~Pradisi, \emph{{Discrete torsion in
  nongeometric orbifolds and their open string descendants}},
  \href{http://dx.doi.org/10.1016/S0550-3213(99)00765-8}{Nucl.Phys. {\bf B573}
  (2000)  314--334},
\href{http://arxiv.org/abs/hep-th/9910228}{{\tt arXiv:hep-th/9910228
  [hep-th]}}.
%%CITATION = HEP-TH/9910228;%%.

\bibitem{Hellerman:2002ax}
S.~Hellerman, J.~McGreevy, and B.~Williams, \emph{{Geometric constructions of
  nongeometric string theories}},
  \href{http://dx.doi.org/10.1088/1126-6708/2004/01/024}{JHEP {\bf 0401} (2004)
   024},
\href{http://arxiv.org/abs/hep-th/0208174}{{\tt arXiv:hep-th/0208174
  [hep-th]}}.
%%CITATION = HEP-TH/0208174;%%.

\bibitem{Hull:2004in}
C.~Hull, \emph{{A Geometry for non-geometric string backgrounds}},
  \href{http://dx.doi.org/10.1088/1126-6708/2005/10/065}{JHEP {\bf 0510} (2005)
   065},
\href{http://arxiv.org/abs/hep-th/0406102}{{\tt arXiv:hep-th/0406102
  [hep-th]}}.
%%CITATION = HEP-TH/0406102;%%.

\bibitem{Flournoy:2004vn}
A.~Flournoy, B.~Wecht, and B.~Williams, \emph{{Constructing nongeometric vacua
  in string theory}},
  \href{http://dx.doi.org/10.1016/j.nuclphysb.2004.11.005}{Nucl.Phys. {\bf
  B706} (2005)  127--149},
\href{http://arxiv.org/abs/hep-th/0404217}{{\tt arXiv:hep-th/0404217
  [hep-th]}}.
%%CITATION = HEP-TH/0404217;%%.

\bibitem{Dabholkar:2005ve}
A.~Dabholkar and C.~Hull, \emph{{Generalised T-duality and non-geometric
  backgrounds}}, \href{http://dx.doi.org/10.1088/1126-6708/2006/05/009}{JHEP
  {\bf 0605} (2006)  009},
\href{http://arxiv.org/abs/hep-th/0512005}{{\tt arXiv:hep-th/0512005
  [hep-th]}}.
%%CITATION = HEP-TH/0512005;%%.

\bibitem{Gray:2005ea}
J.~Gray and E.~J. Hackett-Jones, \emph{{On T-folds, G-structures and
  supersymmetry}}, \href{http://dx.doi.org/10.1088/1126-6708/2006/05/071}{JHEP
  {\bf 0605} (2006)  071},
\href{http://arxiv.org/abs/hep-th/0506092}{{\tt arXiv:hep-th/0506092
  [hep-th]}}.
%%CITATION = HEP-TH/0506092;%%.

\bibitem{Vegh:2008jn}
D.~Vegh and J.~McGreevy, \emph{{Semi-Flatland}},
  \href{http://dx.doi.org/10.1088/1126-6708/2008/10/068}{JHEP {\bf 0810} (2008)
   068},
\href{http://arxiv.org/abs/0808.1569}{{\tt arXiv:0808.1569 [hep-th]}}.
%%CITATION = ARXIV:0808.1569;%%.

\bibitem{Grana:2008yw}
M.~Grana, R.~Minasian, M.~Petrini, and D.~Waldram, \emph{{T-duality,
  Generalized Geometry and Non-Geometric Backgrounds}},
  \href{http://dx.doi.org/10.1088/1126-6708/2009/04/075}{JHEP {\bf 0904} (2009)
   075},
\href{http://arxiv.org/abs/0807.4527}{{\tt arXiv:0807.4527 [hep-th]}}.
%%CITATION = ARXIV:0807.4527;%%.

%\cite{McOrist:2010jw}
\bibitem{McOrist:2010jw}
  J.~McOrist, D.~R.~Morrison and S.~Sethi,
  \emph{Geometries, Non-Geometries, and Fluxes},
  Adv.\ Theor.\ Math.\ Phys.\  {\bf 14} (2010)
\href{http://arxiv.org/abs/arXiv:1004.5447}{{\tt arXiv:1004.5447 [hep-th]}}.
  %%CITATION = ARXIV:1004.5447;%%

%\cite{Hohm:2011zr}
\bibitem{Hohm:2011zr}
  O.~Hohm, S.~K.~Kwak and B.~Zwiebach,
  \emph{Unification of Type II Strings and T-duality},
\href{http://dx.doi.org/10.1103/PhysRevLett.107.171603}{Phys.\ Rev.\ Lett.\  {\bf 107}, 171603 (2011)}
\href{http://arxiv.org/abs/1106.5452}{{\tt arXiv:1106.5452 [hep-th]}}.
  %%CITATION = ARXIV:1106.5452;%%

%\cite{Hohm:2011dv}
\bibitem{Hohm:2011dv}
  O.~Hohm, S.~K.~Kwak and B.~Zwiebach,
  \emph{Double Field Theory of Type II Strings},
\href{http://dx.doi.org/10.1007/JHEP09(2011)013}{JHEP {\bf 1109} (2011) 013}
\href{http://arxiv.org/abs/1107.0008}{{\tt arXiv:1107.0008 [hep-th]}}.
  %%CITATION = ARXIV:1107.0008;%%

\bibitem{Coimbra:2011nw}
A.~Coimbra, C.~Strickland-Constable, and D.~Waldram, \emph{{Supergravity as
  Generalised Geometry I: Type II Theories}},
  \href{http://dx.doi.org/10.1007/JHEP11(2011)091}{JHEP {\bf 1111} (2011)
  091},
\href{http://arxiv.org/abs/1107.1733}{{\tt arXiv:1107.1733 [hep-th]}}.
%%CITATION = ARXIV:1107.1733;%%.


\bibitem{Coimbra:2011ky}
A.~Coimbra, C.~Strickland-Constable, and D.~Waldram, \emph{{$E_{d(d)} \times
  \mathbb{R}^+$ Generalised Geometry, Connections and M theory}},
\href{http://arxiv.org/abs/1112.3989}{{\tt arXiv:1112.3989 [hep-th]}}.
%%CITATION = ARXIV:1112.3989;%%.

\bibitem{Coimbra:2012af}
A.~Coimbra, C.~Strickland-Constable, and D.~Waldram, \emph{{Supergravity as
  Generalised Geometry II: $E_{d(d)} \times \mathbb{R}^+$ and M theory}},
\href{http://arxiv.org/abs/1212.1586}{{\tt arXiv:1212.1586 [hep-th]}}.
%%CITATION = ARXIV:1212.1586;%%.

%\cite{Siegel:1993xq}
\bibitem{Siegel:1993xq}
  W.~Siegel,
  \emph{Two vierbein formalism for string inspired axionic gravity},
\href{http://dx.doi.org/10.1103/PhysRevD.47.5453}{Phys.\ Rev.\ D {\bf 47} (1993) 5453}
\href{http://arxiv.org/abs/hep-th/9302036}{{\tt hep-th/9302036}}.
  %%CITATION = HEP-TH/9302036;%%

%\cite{Siegel:1993th}
\bibitem{Siegel:1993th}
  W.~Siegel,
  \emph{Superspace duality in low-energy superstrings},
\href{http://dx.doi.org/10.1103/PhysRevD.48.2826}{Phys.\ Rev.\ D {\bf 48} (1993) 2826}
\href{http://arxiv.org/abs/hep-th/9305073}{hep-th/9305073}.
  %%CITATION = HEP-TH/9305073;%%


%\cite{West:2010ev}
\bibitem{West:2010ev}
  P.~West,
  \emph{$E_{11}$, generalised space-time and IIA string theory},
\href{http://dx.doi.org/10.1016/j.physletb.2010.12.041}{Phys.\ Lett.\ B {\bf 696} (2011) 403}
\href{http://arxiv.org/abs/arXiv:1009.2624}{{\tt arXiv:1009.2624 [hep-th]}}.
  %%CITATION = ARXIV:1009.2624;%%

%\cite{Rocen:2010bk}
\bibitem{Rocen:2010bk}
  A.~Rocen and P.~West,
  \emph{E11, generalised space-time and IIA string theory: the R-R sector},
in \href{http://dx.doi.org/10.1142/9789814412551_0020}{`Strings, Gauge Fields, and the Geometry Behind',
The Legacy of Maximilian Kreuzer}
\href{http://arxiv.org/abs/arXiv:1012.2744}{{\tt arXiv:1012.2744 [hep-th]}}.
  %%CITATION = ARXIV:1012.2744;%%

%\cite{Andriot:2011uh}
\bibitem{Andriot:2011uh}
  D.~Andriot, M.~Larfors, D.~Lust and P.~Patalong,
  \emph{A ten-dimensional action for non-geometric fluxes},
\href{http://dx.doi.org/10.1007/JHEP09(2011)134}{JHEP {\bf 1109} (2011) 134}
\href{http://arxiv.org/abs/arXiv:1106.4015}{{\tt arXiv:1106.4015 [hep-th]}}.
  %%CITATION = ARXIV:1106.4015;%%

%\cite{Andriot:2013txa}
\bibitem{Andriot:2013txa}
  D.~Andriot,
  \emph{Non-geometric fluxes versus (non)-geometry},
\href{http://arxiv.org/abs/arXiv:1303.0251}{{\tt  arXiv:1303.0251 [hep-th]}}.
  %%CITATION = ARXIV:1303.0251;%%

%\cite{Aldazabal:2013sca}
\bibitem{Aldazabal:2013sca}
  G.~Aldazabal, D.~Marques and C.~Nunez,
  \emph{Double Field Theory: A Pedagogical Review},
\href{http://dx.doi.org/10.1088/0264-9381/30/16/163001}{Class.\ Quant.\ Grav.\  {\bf 30} (2013) 163001}
\href{http://arxiv.org/abs/arXiv:1305.1907}{{\tt arXiv:1305.1907 [hep-th]}}.
  %%CITATION = ARXIV:1305.1907;%%


\bibitem{Grana:2001xn}
M.~Grana and J.~Polchinski, \emph{{Gauge / gravity duals with holomorphic
  dilaton}}, \href{http://dx.doi.org/10.1103/PhysRevD.65.126005}{Phys.Rev. {\bf
  D65} (2002)  126005},
\href{http://arxiv.org/abs/hep-th/0106014}{{\tt arXiv:hep-th/0106014
  [hep-th]}}.
%%CITATION = HEP-TH/0106014;%%.

\bibitem{kumarelkies}
N.~Elkies and A.~Kumar, \emph{{ K3 surfaces and equations for Hilbert modular
  surfaces }}, \href{http://arxiv.org/abs/arXiv:1209.3527}{{\tt
  arXiv:1209.3527}}.

\bibitem{LopesCardoso:1996hq}
G.~Lopes~Cardoso, G.~Curio, D.~Lust, and T.~Mohaupt, \emph{{On the duality
  between the heterotic string and F theory in eight-dimensions}},
  \href{http://dx.doi.org/10.1016/S0370-2693(96)01303-2}{Phys.Lett. {\bf B389}
  (1996)  479--484},
\href{http://arxiv.org/abs/hep-th/9609111}{{\tt arXiv:hep-th/9609111
  [hep-th]}}.
%%CITATION = HEP-TH/9609111;%%.

\bibitem{Morrison:1996pp}
D.~R. Morrison and C.~Vafa, \emph{{Compactifications of F theory on Calabi-Yau
  threefolds. 2.}},
  \href{http://dx.doi.org/10.1016/0550-3213(96)00369-0}{Nucl.Phys. {\bf B476}
  (1996)  437--469},
\href{http://arxiv.org/abs/hep-th/9603161}{{\tt arXiv:hep-th/9603161
  [hep-th]}}.
%%CITATION = HEP-TH/9603161;%%.

\bibitem{M1}
D. R. Morrison, \emph{On K3 surfaces with large Picard number}, Invent. Math. 75
(1984) no. 1, 105–121.


\bibitem{AKumarg2}
A.~Kumar, \emph{{K3 surfaces associated to curves of genus two }},
  \href{http://arxiv.org/abs/arXiv:math/0701669}{{\tt arXiv:math/0701669}}.


\bibitem{Sen:1997gv}
A.~Sen, \emph{{Orientifold limit of F theory vacua}},
  \href{http://dx.doi.org/10.1103/PhysRevD.55.R7345}{Phys.Rev. {\bf D55} (1997)
   7345--7349},
\href{http://arxiv.org/abs/hep-th/9702165}{{\tt arXiv:hep-th/9702165
  [hep-th]}}.
%%CITATION = HEP-TH/9702165;%%.




\bibitem{peters}
W.~Barth, C.~Peters, and V.~de~Ven, \emph{{Compact complex surfaces }}, Ergeb.
  Math. Grenzgeb. (3) 4, Springer-Verlag, Berlin, 1984  .

\bibitem{schuettshioda}
M.~Schuett and T.~Shioda, \emph{{Elliptic Surfaces }},
  \href{http://arxiv.org/abs/0907.0298 [math.AG]}{{\tt arXiv:0907.0298
  [math.AG]}}.

\bibitem{Aspinwall:1996mn}
P.~S. Aspinwall, \emph{{K3 surfaces and string duality}},
\href{http://arxiv.org/abs/hep-th/9611137}{{\tt arXiv:hep-th/9611137
  [hep-th]}}.
%%CITATION = HEP-TH/9611137;%%.

\bibitem{GH}
P.Griffith and J.Harris, {\em Principles of Algebraic Geometry}.
\newblock Wiley-Interscience, New York, 1978.

\bibitem{Denef:2008wq}
F.~Denef, \emph{{Les Houches Lectures on Constructing String Vacua}},
\href{http://arxiv.org/abs/0803.1194}{{\tt arXiv:0803.1194 [hep-th]}}.
%%CITATION = ARXIV:0803.1194;%%.


\bibitem{arnold}
V.~I. Arnold, S.~M. Gusein-Zade, and A.~N. Varchenko, {\em Singularities of
  Differentiable Maps}.
\newblock Birkh\"auser, Boston, 1988.

\bibitem{Aspinwall:1997eh}
P.~S. Aspinwall, \emph{{M theory versus F theory pictures of the heterotic
  string}}, Adv.Theor.Math.Phys. {\bf 1} (1998)  127--147,
\href{http://arxiv.org/abs/hep-th/9707014}{{\tt arXiv:hep-th/9707014
  [hep-th]}}.
%%CITATION = HEP-TH/9707014;%%.

\bibitem{Gaberdiel:1997ud}
M.~R. Gaberdiel and B.~Zwiebach, \emph{{Exceptional groups from open strings}},
  \href{http://dx.doi.org/10.1016/S0550-3213(97)00841-9}{Nucl.Phys. {\bf B518}
  (1998)  151--172},
\href{http://arxiv.org/abs/hep-th/9709013}{{\tt arXiv:hep-th/9709013}}.
%%CITATION = HEP-TH/9709013;%%.

%\cite{DeWolfe:1998pr}
\bibitem{DeWolfe:1998pr}
  O.~DeWolfe, T.~Hauer, A.~Iqbal and B.~Zwiebach,
  \emph{Uncovering infinite symmetries on [p, q] 7-branes: Kac-Moody algebras and beyond},
  Adv.\ Theor.\ Math.\ Phys.\  {\bf 3} (1999) 1835
\href{http://arxiv.org/abs/hep-th/9812209}{{\tt arXiv:hep-th/9812209}}.
  %%CITATION = HEP-TH/9812209;%%

\bibitem{Hayashi:2010zp}
H.~Hayashi, T.~Kawano, Y.~Tsuchiya, and T.~Watari, \emph{{More on Dimension-4
  Proton Decay Problem in F-theory -- Spectral Surface, Discriminant Locus and
  Monodromy}},
  \href{http://dx.doi.org/10.1016/j.nuclphysb.2010.07.011}{Nucl.Phys. {\bf
  B840} (2010)  304--348},
\href{http://arxiv.org/abs/1004.3870}{{\tt arXiv:1004.3870 [hep-th]}}.
%%CITATION = ARXIV:1004.3870;%%.

\bibitem{Braun:2008ua}
A.~Braun, A.~Hebecker, and H.~Triendl, \emph{{D7-Brane Motion from M-Theory
  Cycles and Obstructions in the Weak Coupling Limit}},
  \href{http://dx.doi.org/10.1016/j.nuclphysb.2008.03.021}{Nucl.Phys. {\bf
  B800} (2008)  298--329},
\href{http://arxiv.org/abs/0801.2163}{{\tt arXiv:0801.2163 [hep-th]}}.
%%CITATION = ARXIV:0801.2163;%%.

\end{thebibliography}
\end{document}